\documentclass[pra,twocolumn,balancelastpage,letterpaper]{revtex4}

\usepackage{amsmath}
\usepackage{amssymb}
\usepackage{mathrsfs}
\usepackage{xspace}
\usepackage{graphicx}
\usepackage{braket}
\usepackage{slashed}
\usepackage[utf8]{inputenc}
\usepackage{flushend}

\usepackage{ifpdf}
\ifpdf
\fi

\setcitestyle{numbers,square}
\usepackage{hyperref}

\hypersetup{
  colorlinks   = true, 
  urlcolor     = blue, 
  linkcolor    = blue, 
  citecolor   = blue 
}
       
\newcommand{\eq}[1]{(\ref{#1})}
\newcommand{\Eq}[1]{Eq.~(\ref{#1})}
\newcommand{\Eqs}[1]{Eqs.~(\ref{#1})}

\newcommand{\Sec}[1]{Sec.~\ref{#1}}
\newcommand{\Ref}[1]{Ref.~\cite{#1}}
\newcommand{\Refs}[1]{Refs.~\cite{#1}}
\newcommand{\App}[1]{Appendix~\ref{#1}}

\newcommand{\eg}{{e.g.,\/}\xspace}
\newcommand{\ie}{{i.e.,\/}\xspace}

\newcommand{\pd}{\partial}
\newcommand{\del }{\vec{\nabla}}
\newcommand{\cc}{\text{c.\,c.}}
\newcommand{\hc}{\text{h.\,c.}}

\newcommand{\avg}[1]{\left\langle #1 \right\rangle }
\newcommand{\mc}[1]{\mathcal{#1}}
\newcommand{\mcc}[1]{\mathfrak{#1}}
\newcommand{\msf}[1]{\mathsf{#1}}

\renewcommand{\vec}[1]{{\boldsymbol{\rm #1}}}

\newcommand{\pik}{(\msf{\pi} \cdot \msf{k})} 
\newcommand{\Pik}{(\msf{\Pi} \cdot \msf{k})} 
\newcommand{\sla}[1]{\slashed{#1}}

\usepackage{color} 
\usepackage{blindtext}

\begin{document}

\title{Relativistic ponderomotive Hamiltonian of a Dirac particle in a vacuum laser field}

\begin{abstract}

We report a point-particle ponderomotive model of a Dirac electron oscillating in a high-frequency field. Starting from the Dirac Lagrangian density, we derive a reduced phase-space Lagrangian that describes the relativistic time-averaged dynamics of such a particle in a geometrical-optics laser pulse propagating in vacuum. The pulse is allowed to have an arbitrarily large amplitude provided that radiation damping and pair production are negligible. The model captures the Bargmann-Michel-Telegdi (BMT) spin dynamics, the Stern-Gerlach spin-orbital coupling, the conventional ponderomotive forces, and the interaction with large-scale background fields (if any). Agreement with the BMT spin precession equation is shown numerically. The commonly known theory in which ponderomotive effects are incorporated in the particle effective mass is reproduced as a special case when the spin-orbital coupling is negligible. This model could be useful for studying laser-plasma interactions in relativistic spin-$1/2$ plasmas.

\end{abstract}

\author{D.~E. Ruiz, C.~L.~Ellison, and I.~Y. Dodin}
\affiliation{Department of Astrophysical Sciences, Princeton University, Princeton, New Jersey 08544, USA}
\date{\today}

\maketitle

\bibliographystyle{apsrev-title}

\section{Introduction}

In recent years, many works have been focused on incorporating quantum effects into classical plasma dynamics \cite{Melrose:2008bw,Brodin:2008cl}. In particular, various models have been proposed to marry spin equations with classical equations of plasma physics. This includes the early works by Takabayasi \cite{Takabayasi:1955iy,Takabayasi:1957tz} as well as the most recent works presented in \Refs{Marklund:2007cv,Brodin:2008er,Brodin:2011kf,Stefan:2013hr,Dixit:2013ff, Morandi:2014jm,Andreev:1746689,Andreev:2015if,Andreev:2014uw}. Of particular interest in this regard is the regime wherein particles interact with high-frequency electromagnetic (EM) radiation. In this regime, it is possible to introduce a simpler time-averaged description, in which particles experience effective time-averaged, or ``ponderomotive," forces \cite{Boot:1957um,Gaponov:1958un, Cary:1977ud}. It was shown recently that the inclusion of spin effects yields intriguing corrections to this time-averaged dynamics  \cite{Brodin:2010bc,Stefan:2011cq}. However, current ``spin-ponderomotive''  theories remain limited to regimes where (i) the particle de~Broglie wavelength is much less than the radiation wavelength and (ii) the radiation amplitude is small enough so that it can be treated as a perturbation. These conditions are far more restrictive than those of spinless particle theories, where non-perturbative, relativistic ponderomotive effects can be accommodated within the effectively modified particle mass \cite{Akhiezer:1956wb,Kibble:1966fj,Kibble:1966eb,Raicher:2014gj, foot:dodin}. One may wonder then: is it possible to derive a fully relativistic, and yet transparent, theory accounting also for the spin dynamics and the Stern-Gerlach-type spin-orbital coupling?

Excitingly, the answer is yes, and the purpose of this paper is to propose such a description. More specifically, what we report here is a point-particle ponderomotive model of a Dirac electron \cite{Dirac:1928dh}. Starting from the Dirac Lagrangian density, we derive a phase-space Lagrangian \eq{eq:lagr_point_pond} in canonical coordinates with a Hamiltonian \eq{eq:H_point_pond} that describes the relativistic time-averaged dynamics of such particle in a geometrical-optics (GO) laser pulse propagating in vacuum \cite{Tracy:2014to}. The pulse is allowed to have an arbitrarily large amplitude (as long as radiation damping and pair production are negligible) and, in case of nonrelativistic interactions, a wavelength comparable to the electron de Broglie wavelength. The model captures the spin dynamics, the spin-orbital coupling, the conventional ponderomotive forces, and the interaction with large-scale background fields (if any). Agreement with the Bargmann-Michel-Telegdi (BMT) spin precession equation \cite{Bargmann:1959us} is shown numerically. The aforementioned ``effective-mass'' theory for spinless particles is reproduced as a special case when the spin-orbital coupling is negligible. Also notably, the point-particle Lagrangian that we derive has a canonical structure, which could be helpful in simulating the corresponding dynamics using symplectic methods \cite{Hairer_2006,McLachlan_2006,Qin_2015}.

This work is organized as follows. In \Sec{sec:notation} the basic notation is defined. In \Sec{sec:basic} the main assumptions used throughout the work are presented. To arrive at the point-particle ponderomotive model, Secs. \ref{sec:pondero}-\ref{sec:point} apply successive approximations and reparameterizations to the Dirac Lagrangian density. Specifically, in \Sec{sec:pondero} we derive a ponderomotive Lagrangian density that captures the average dynamics of a Dirac particle. In \Sec{sec:reduced} we obtain a reduced Lagrangian model that explicitly shows orbital-spin coupling effects. In \Sec{sec:wave} we deduce a ``fluid" Lagrangian model that describes the particle wave packet dynamics. In \Sec{sec:point} we calculate the point-particle limit of such ``fluid" model. In \Sec{sec:numerical} the ponderomotive model is numerically compared to a generalized non-averaged BMT model. In \Sec{sec:conclusions} the main results are summarized.


\section{Notation}
\label{sec:notation}

The following notation is used throughout the paper. The symbol ``$\doteq$'' denotes definitions, ``\hc'' denotes ``Hermitian conjugate,'' and ``\cc '' denotes ``complex conjugate.'' Unless indicated otherwise, we use natural units so that the speed of light and the Plank constant equal unity ($c = \hbar = 1$). The identity $N\times N$ matrix is denoted by $\mathbb{I}_{N}$. The Minkowski metric is adopted with signature $(+, -, -, -)$. Greek indices span from $0$ to $3$ and refer to spacetime coordinates $x^\mu=(x^0, \vec{x})$ with $x^0$ corresponding to the time variable $t$. Also, $\pd_\mu \equiv \pd/\pd x^\mu=(\pd_t, \del)$ and $\mathrm{d}^4x \equiv \mathrm{d}t\,\mathrm{d}^3x$. Latin indices span from $1$ to $3$ and denote the spatial variables, \ie $\vec{x} = (x^1, x^2, x^3)$ and $\pd_i \equiv \pd/\pd x^i$. Summation over repeated indexes is assumed. In particular, for arbitrary four-vectors $\msf{a}$ and $\msf{b}$, we have $\msf{a}\cdot \msf{b} \equiv a^\mu b_\mu = a^0 b^0 - \vec{a}\cdot \vec{b}$ and $\msf{a}^2 \equiv \msf{a} \cdot \msf{a}$. The Feynman slash notation is used: $\sla{a} \doteq a_\mu \gamma^\mu$, where $\gamma^\mu=(\gamma^0, \vec{\gamma})$ are the Dirac matrices (see below). The average of an arbitary complex-valued function $f(\msf{x},\Theta)$ with respect to a phase $\Theta$ is denoted by $\avg{f}$. In Euler-Lagrange equations (ELEs), the notation ``$\delta a: $'' means that the corresponding equation is obtained by extremizing the action integral with respect to $a$.

\section{Basic Formalism}
\label{sec:basic}

As for any quantum particle or non-dissipative wave \cite{Dodin:2014hw}, the dynamics of a Dirac electron \cite{Dirac:1928dh} is governed by the least action principle $\, \delta \mc{A} \,= \, 0$, where $\mc{A}$ is the action
\begin{gather}\label{eq:action}
\mc{A} = \int \mcc{L}\,\mathrm{d}^4x,
\end{gather}
and $\mcc{L}$ is the Lagrangian density given by \cite{foot:anomalous}
\begin{equation}
\mcc{L}= \frac{i}{2} \left[ \bar{\psi} \gamma^\mu (\pd_\mu \psi) -( \pd_\mu \bar{\psi}) \gamma^\mu \psi \right] 
	 - \bar{\psi}  q \sla{A} \psi - \bar{\psi} m  \psi .
\label{eq:lagr}
\end{equation}
Here $q$ and $m$ are the particle charge and mass, $\psi$ is a complex four-component wave function, and $\bar{\psi} \doteq \psi^\dag\gamma^0$ is its Dirac conjugate. The Dirac matrices $\gamma^\mu$ satisfy
\begin{equation}
\gamma^\mu \gamma^\nu + \gamma^\nu \gamma^\mu = 2 g^{\mu \nu} \mathbb{I}_4,
\label{eq:anticommutation}
\end{equation}
where $g^{\mu \nu}$ is the Minkowski metric tensor. Hence,
\begin{gather}
\sla{a}	\sla{b}  + \sla{b} \sla{a} = 2 (\msf{a} \cdot \msf{b}) \mathbb{I}_4 ,	\label{eq:slashed_1}		\\
\sla{a} \sla{a}= \msf{a}^2 \mathbb{I}_4,		\label{eq:slashed_2}
\end{gather}
for any pair of four-vectors $\msf{a}$ and $\msf{b}$. In this work, the standard representation of the Dirac matrices is used:
\begin{equation}
\gamma^0 =
\begin{pmatrix}
	\mathbb{I}_2 & 0 \\
	0 & -\mathbb{I}_2
\end{pmatrix},
\quad
\vec{\gamma} =
	\begin{pmatrix}
	0 & \vec{\sigma} \\
	-\vec{\sigma} & 0 \\
\end{pmatrix}   ,
\end{equation}
where $\vec{\sigma}=( \sigma_x, \sigma_y, \sigma_z)$ are the $2\times 2$ Pauli matrices. Note that these matrices satisfy
\begin{equation}
(\gamma^\mu)^\dag = \gamma^0 \gamma^\mu \gamma^0.
\label{eq:gamma_dag}
\end{equation}

We consider the interaction of an electron with an EM field such that the four-vector potential $\msf{A}$ has the form
\begin{equation}
\msf{A}(\epsilon \msf{x}, \Theta)= \msf{A}_{\rm bg}(\epsilon \msf{x})	+		\msf{A}_{\rm osc}(\epsilon \msf{x}, \Theta).
\end{equation}
Here $\msf{A}_{\rm bg}(\epsilon \msf{x})$ describes a background field (if any) that is slow, as determined by a small dimensionless parameter $\epsilon$, which is yet to be specified. The other part of the vector potential,
\begin{equation}
\msf{A}_{\rm osc}(\epsilon \msf{x}, \Theta) = 
			\mathrm{Re} 	\left[		\msf{A}_{\rm osc,c} (\epsilon \msf{x}) e^{i\Theta}	\right],
\end{equation}
describes a rapidly oscillating EM wave field, \eg a laser pulse.  Here $\msf{A}_{\rm osc,c} (\epsilon \msf{x})$ is a complex four-vector describing the laser envelope with a slow spacetime dependence, and $\Theta$ is a rapid phase. The EM wave frequency is defined by $\omega(\epsilon \msf{x})\doteq -\pd_t \Theta$, and the wave vector is $\vec{k}(\epsilon \msf{x})\doteq \del \Theta$. Accordingly, $k^\mu \doteq -\pd^\mu \Theta = (\omega, \vec{k})$. We describe $\msf{A}_{\rm osc}$ within the GO approximation \cite{Dodin:2012hn} and assume that the interaction takes place in vacuum. Then, the four-wavevector $\msf{k}$ satisfies the vacuum dispersion relation
\begin{equation}\label{eq:k_dispersion}
	\msf{k}^2 = \omega^2 - \vec{k}^2=0, 
\end{equation}
which can also be expressed as
\begin{gather}\label{eq:KK}
	\sla{k}\sla{k} = \msf{k}^2 \mathbb{I}_4 = 0.
\end{gather}
Furthermore, a Lorentz gauge condition is chosen for the oscillatory field such that 
\begin{equation}
\pd_\mu A_{\rm osc}^\mu = 0.
\label{eq:lorentz}
\end{equation}

In this work, we neglect radiation damping and assume
\begin{equation}
		\omega'/\omega_c \ll 1,
\label{eq:ordering1}
\end{equation}
where $\omega_{\rm c}\doteq m $ is the Compton frequency and $\omega'$ is the frequency in the electron rest frame. Then, pair production (and annihilation) can be neglected. We also assume
\begin{equation}
\epsilon \doteq \mathrm{max} \left( \frac{1}{\omega \tau}	,	\frac{1}{|\vec{k}| \ell } 	\right) 		\ll 1, 
\label{eq:ordering2}
\end{equation}
where $\tau$ and $\ell$ are the characteristic temporal and spatial scales of $\msf{k}$, $\msf{A}_{\rm bg}$, and $\msf{A}_{\rm osc,c}$. Using this ordering and the Lagrangian density \eq{eq:lagr}, we aim to derive a reduced Lagrangian density that describes the ponderomotive ($\Theta$-averaged) dynamics of an electron accurately enough to capture the spin-orbital coupling effects to the leading order in $\epsilon$. As shown in \Refs{Ruiz:2015hq, Ruiz:2015bz}, this requires that $\mc{O}(\epsilon)$ terms be retained when approximating the Lagrangian density \eq{eq:lagr}. Such reduced Lagrangian density is derived as follows.

\section{Ponderomotive Model}
\label{sec:pondero}

In this section, we derive a ponderomotive Lagrangian density for the four-component Dirac wave function.

\subsection{Wave function parameterization}
\label{sec:parameterization}

Consider the following representation for the four-component wave function:
\begin{equation}
\psi(\msf{x}) = \xi e^{i \theta}.
\end{equation}
Here $\theta(\msf{x})$ is a fast real phase, and $\xi(\epsilon \msf{x}, \Theta)$ is a complex four-component vector slow compared to $\theta(\msf{x})$. In these variables, the Lagrangian density \eq{eq:lagr} is expressed as

\begin{multline}
\mcc{L} = \frac{i}{2} \left[ \bar{\xi} \gamma^\mu (\pd_\mu \xi) -( \pd_\mu \bar{\xi}) \gamma^\mu \xi \right] \\
		+ \bar{\xi}  \left( \sla{\pi}  -q \sla{A}_{\rm osc}  -  m \mathbb{I}_4 \right) \xi ,
\label{eq:lagr_comparison}
\end{multline}
where 
\begin{gather}
	\pi^\mu(\epsilon \msf{x} ) \doteq p^\mu - q A_{\rm bg}^\mu, \\
	p_\mu(\epsilon \msf{x} ) \doteq -\pd_\mu \theta .
\end{gather}

It is convenient to parameterize $\xi$ in terms of the ``semiclassical'' Volkov solution (\App{app:volkov}) since the latter becomes the exact solution of the Dirac equation in the limit of vanishing $\epsilon$. Specifically, we write
\begin{equation}
\xi(\epsilon \msf{x}, \Theta)= \Xi e^{i \tilde{\theta}}  \varphi .
\label{eq:volkov}
\end{equation}
Here $\varphi (\epsilon \mathsf{x}, \Theta)$ is a near-constant function with an asymptotic representation of the form
\begin{equation}
\varphi( \epsilon \msf{x}, \Theta) = \sum_{n=-\infty}^\infty \epsilon^{|n|} \varphi_n(\epsilon \msf{x}) e^{i n\Theta}.
\label{eq:asymptotic} 
\end{equation}
(so that $\varphi \to \varphi_0 \to \text{const}$ at $\epsilon \to 0$), the real phase $\tilde{\theta}(\epsilon \mathsf{x}, \Theta)$ is given by
\begin{multline}
\tilde{\theta}(\epsilon \msf{x}, \Theta) \doteq \frac{q}{\msf{\pi} \cdot \msf{k} } \int^\Theta  \msf{\pi} \cdot \msf{A}_{\rm osc} \, \mathrm{d}\Theta' \\
-\frac{q^2}{2 \pik } \int^\Theta 
 \left( \msf{A}_{\rm osc}^2 - \avg{\msf{A}_{\rm osc}^2 } \right)  \, \mathrm{d}\Theta'
 \label{eq:Ip}
\end{multline}
and has the property $\langle \tilde{\theta} \rangle = 0$, and $\Xi(\epsilon \mathsf{x}, \Theta)$ is a matrix defined as follows:
\begin{equation}
\Xi(\epsilon \msf{x}, \Theta) \doteq 
		 \mathbb{I}_4 + \frac{q}{2 \pik} \sla{k} \sla{A}_{\rm osc} .
					\label{eq:Xi} 
\end{equation}
Notice also that the Dirac conjugate of $\xi$ is given by
\begin{equation}
\bar{\xi} =(\Xi e^{i \tilde{\theta}} \varphi  )^\dag \gamma^0 = e^{-i \tilde{\theta} } \varphi^\dag   \Xi^\dag \gamma^0 = e^{-i \tilde{\theta} } \bar{\varphi} \gamma^0 \Xi^\dag \gamma^0,
\label{eq:xi_bar}
\end{equation}
where 
\begin{align}
\gamma^0 \Xi^\dag \gamma^0 
			=&   \gamma^0  \left[ \mathbb{I}_4+ \frac{q}{2 \pik } \sla{A}^\dag_{\rm osc} \gamma^0 \gamma^0\sla{k}^\dag \right] \gamma^0 \notag \\
			=&   \mathbb{I}_4+ \frac{q}{2 \pik } \sla{A}_{\rm osc} \sla{k} .
\label{eq:xi_special}
\end{align}
Here we used \Eqs{eq:anticommutation}, \eq{eq:gamma_dag}, and \eq{eq:Xi}. 


\subsection{Lagrangian density in the new variables}

Inserting \Eqs{eq:volkov} and \eq{eq:xi_bar} into \Eq{eq:lagr_comparison} leads to
\begin{align}
\mcc{L} =&
		 \underbrace{ \bar{\varphi}  \gamma^0 \Xi^\dag \gamma^0 
		 						\sla{\pi} \Xi \varphi  }_{=\mcc{L}_\textrm{1} }  
		\underbrace{ - \bar{\varphi}  m \gamma^0 \Xi^\dag \gamma^0  
								\Xi \varphi  }_{=\mcc{L}_\textrm{2}} 
		\underbrace{ -\bar{\varphi}  \gamma^0 \Xi^\dag \gamma^0 q\sla{A}_{\rm osc} 
								\Xi \varphi }_{=\mcc{L}_\textrm{3}}   \notag \\
 &		+\underbrace{ \frac{i}{2}\left[ \bar{\varphi} \gamma^0 \Xi^\dag \gamma^0\gamma^\mu 			
						\Xi (\pd_\mu \varphi) - \cc	\right] }_{=\mcc{L}_\textrm{4}}  				
 		\underbrace{ -\bar{\varphi}  \gamma^0 \Xi^\dag \gamma^0 (\sla{\pd} \tilde{\theta} )\Xi \varphi
 							}_{=\mcc{L}_\textrm{5}}  				\notag \\
 &		+\underbrace{ \frac{i}{2} \bar{\varphi}  \left[  \gamma^0 \Xi^\dag \gamma^0
					  \gamma^\mu (\pd_\mu \Xi)- \hc \right] \varphi  } _{=\mcc{L}_\textrm{6}}.
\label{eq:lagr_trans}
\end{align}
Let us explicitly calculate each term in \Eq{eq:lagr_trans}. Substituting \Eqs{eq:Xi} and \eq{eq:xi_special} into $\mcc{L}_1$ leads to
\begin{align}
\mcc{L}_1  =&  \bar{\varphi} \gamma^0 \Xi^\dag \gamma^0 \sla{\pi} \Xi   \varphi \notag \\
					=&  \bar{\varphi}  \left[ \mathbb{I}_4+ \frac{q}{2  \pik } \sla{A}_{\rm osc} \sla{k} \right]  \sla{\pi} 
												\left[ \mathbb{I}_4+ \frac{q}{2 \pik } \sla{k} \sla{A}_{\rm osc} \right]   \varphi  \notag \\
					=&  \bar{\varphi}  \left[ \sla{\pi}  
						+ \frac{q}{2  \pik } ( \sla{A}_{\rm osc}  \sla{k} \sla{\pi}	+  \sla{\pi} \sla{k} \sla{A}_{\rm osc} )	\right.	\notag \\
					&~~~~~~~~~\left.	
							+ \frac{q^2}{4  \pik^2 }  \sla{A}_{\rm osc} \sla{k}  \sla{\pi} \sla{k} \sla{A}_{\rm osc} 												 \right]  \varphi \notag \\
					=& \bar{\varphi}	\left[	\sla{\pi} 
							+ q\sla{A}_{\rm osc}
							+ \frac{q\msf{A}_{\rm osc} \cdot \msf{k}}{\msf{\pi} \cdot \msf{k} } \sla{\pi}
							- \frac{  q\msf{A}_{\rm osc} \cdot \pi }{\msf{\pi} \cdot \msf{k} } \sla{k}	\right. 	\notag \\
					&~~~~~~~~~\left.		+ \frac{q^2 \msf{A}_{\rm osc} \cdot \msf{k}}{\msf{\pi} \cdot \msf{k} } \sla{A}_{\rm osc}
							- \frac{q^2 \msf{A}_{\rm osc}^2 }{2\pik} \sla{k}
							\right] \varphi,
\label{eq:term_1}
\end{align}
where we used \Eqs{eq:slashed_1}, \eq{eq:slashed_2}, and \eq{eq:KK}. Similarly,
\begin{align}
\mcc{L}_2   
	=& -	\bar{\varphi} m  \gamma^0 \Xi^\dag \gamma^0  \Xi \varphi \notag \\
	=& -	\bar{\varphi} m   \left[ \mathbb{I}_4+  \frac{q}{2 \pik} \sla{A}_{\rm osc}  \sla{k} \right] 	
								 \left[ \mathbb{I}_4+ \frac{q}{2 \pik} \sla{k} \sla{A}_{\rm osc} \right]  
			  \varphi \notag \\
	=& -	\bar{\varphi} m \left[ \mathbb{I}_4 
					+  \frac{q}{2  \pik}  ( \sla{A}_{\rm osc} \sla{k} + \sla{k} \sla{A}_{\rm osc}    ) \right] \varphi \notag \\
	=& -	\bar{\varphi} m \left(1
						+  \frac{q \msf{A}_{\rm osc} \cdot \msf{k} }{ \msf{\pi} \cdot \msf{k} }  \right) \varphi ,
\label{eq:term_2}
\end{align}
where we used \Eq{eq:KK} to obtain the third line and \Eq{eq:slashed_1} to get the last line. For $\mcc{L}_\textrm{3} $, one obtains
\begin{align}
\mcc{L}_3 
	=& -	\bar{\varphi} \gamma^0 \Xi^\dag \gamma^0 q\sla{A}_{\rm osc} \Xi   \varphi  \notag \\
	=& -	\bar{\varphi} \left[ \mathbb{I}_4+ \frac{q\sla{A}_{\rm osc}  \sla{k}}{2 \pik }  \right] q\sla{A}_{\rm osc}  
						\left[ \mathbb{I}_4+ \frac{q \sla{k} \sla{A}_{\rm osc} }{2 \pik} \right] \varphi \notag \\
	=& -	\bar{\varphi} \left[ 
					q\sla{A}_{\rm osc}  
				+	\frac{q^2}{ \msf{\pi} \cdot \msf{k} } \sla{A}_{\rm osc}  \sla{k}\sla{A}_{\rm osc}  \right.	\notag\\
	&~~~~~~~~\left.		+	\frac{q^3}{ 4 \pik^2 }  
					\sla{A}_{\rm osc}  \sla{k}\sla{A}_{\rm osc}   \sla{k}\sla{A}_{\rm osc}  	\right] \varphi		\notag \\
	=& -	\bar{\varphi} \left[ 
					q\sla{A}_{\rm osc}  
				+	\frac{2q^2 \msf{A}_{\rm osc} \cdot \msf{k} }{ \msf{\pi} \cdot \msf{k} } \sla{A}_{\rm osc}
				-	\frac{q^2 \msf{A}_{\rm osc}^2 }{ \msf{\pi} \cdot \msf{k} } \sla{k} \right.	\notag\\
	&~~~~~~~~\left.		+	\frac{q^3 \msf{A}_{\rm osc} \cdot \msf{k}  }{ 2 \pik^2 }  
										\sla{A}_{\rm osc}  \sla{k}\sla{A}_{\rm osc}    	\right] \varphi.
\label{eq:term_3}
\end{align}

The terms $\mcc{L}_4$, $\mcc{L}_5$, and $\mcc{L}_6$ in \Eq{eq:lagr_trans} involve spacetime derivatives of $(\tilde{\theta}$, $\Xi$, $\varphi)$, which have slow spacetime and rapid $\Theta$ dependences. For convenience, let us write the derivative operator $\pd_\mu$ as follows:
\begin{equation}
\pd_\mu f(\epsilon \msf{x}, \Theta) = \epsilon d_\mu f(\epsilon \msf{x}, \Theta) 
					-k_\mu \pd_\Theta f(\epsilon \msf{x}, \Theta),
\label{eq:notation_der}
\end{equation}
where $f$ is an arbitrary function and $d_\mu$ indicates a derivation with respect to the first argument of $f$. In this notation, $\mcc{L}_4$ can be written as follows:
\begin{align}
\mcc{L}_4   
		&		= \frac{i}{2}\left[ \bar{\varphi} \gamma^0 \Xi^\dag \gamma^0	\gamma^\mu 			
						\Xi (\pd_\mu \varphi) - \cc	\right]		\notag 		\\
		&		= -\frac{i}{2}\left[ \bar{\varphi} \gamma^0 \Xi^\dag \gamma^0	\sla{k} 			
						\Xi (\pd_\Theta \varphi) - \cc	\right]		\notag 	\\
		&~~~~~~~~~~~
					+	\frac{i\epsilon }{2}\left[ \bar{\varphi} \gamma^0 \Xi^\dag \gamma^0	\gamma^\mu 		\Xi (d_\mu \varphi) - \cc	\right]	
						\notag 		\\
		&		= -\frac{i}{2}\left[ \bar{\varphi} 	\sla{k} 	(\pd_\Theta \varphi) - (\pd_\Theta \bar{\varphi}) \sla{k} \varphi	\right]		\notag 	\\
		&~~~~~~~~~~~
					+	\frac{i\epsilon }{2}\left[ \bar{\varphi} \gamma^0 \Xi^\dag \gamma^0	\gamma^\mu 		\Xi (d_\mu \varphi) - \cc	\right],
\label{eq:term_4}
\end{align}
where in the third line, we used \Eqs{eq:KK}, \eq{eq:Xi}, and \eq{eq:xi_special}. Similarly, substituting \Eq{eq:Ip} into $\mcc{L}_5$ leads to
\begin{align}
\mcc{L}_5  
		&		= -  \bar{\varphi} \gamma^0 \Xi^\dag \gamma^0	( \sla{\pd} \tilde{\theta} ) \Xi  \varphi 	\notag 		\\
		&		=   \bar{\varphi} \left[ \mathbb{I}_4+ \frac{q \sla{A}_{\rm osc}  \sla{k} }{2 \pik } \right] \sla{k} (\pd_\Theta \tilde{\theta} )
											\left[ \mathbb{I}_4+ \frac{q \sla{k} \sla{A}_{\rm osc} }{2 \pik} \right] \varphi \notag \\
		&~~~~~ 
					-\epsilon \bar{\varphi} \left[ \mathbb{I}_4+ \frac{q \sla{A}_{\rm osc}  \sla{k} }{2 \pik } \right] (\sla{d} \tilde{\theta} )
											\left[ \mathbb{I}_4+ \frac{q \sla{k} \sla{A}_{\rm osc} }{2 \pik} \right] \varphi \notag 		\\
		&		=   \bar{\varphi} \sla{k} \varphi 
				\left[ \frac{  q\msf{A}_{\rm osc} \cdot \msf{\pi} }{ \msf{\pi} \cdot \msf{k}  } 
							- 	\frac{q^2  \msf{A}_{\rm osc}^2}{2 \pik}
							+	\frac{q^2  \avg{ \msf{A}_{\rm osc}^2 } }{2 \pik}
							\right] \notag \\
		&~~~~~ 
					-\epsilon \bar{\varphi} \left[ \mathbb{I}_4+ \frac{q \sla{A}_{\rm osc}  \sla{k} }{2 \pik } \right] (\sla{d} \tilde{\theta} )
											\left[ \mathbb{I}_4+ \frac{q \sla{k} \sla{A}_{\rm osc} }{2 \pik} \right] \varphi .
\label{eq:term_5}
\end{align}
Finally, the last term $\mcc{L}_6$ gives
\begin{align}
\mcc{L}_6 	&	= \frac{i}{2} \bar{\varphi}  \left[  \gamma^0 \Xi^\dag \gamma^0
								  \gamma^\mu (\pd_\mu \Xi)- \hc \right] \varphi  	\notag \\
				&	= \frac{i\epsilon }{2} \bar{\varphi} \left\lbrace
							\left[ \mathbb{I}_4+ \frac{q \sla{A}_{\rm osc}  \sla{k} }{2 \pik } \right] 
								  \sla{d} \left[ \frac{q \sla{k} \sla{A}_{\rm osc} }{2 \pik} \right]			\right.	\notag	\\
				&~~~~~
							-	\left. 	d_\mu 	\left[ \frac{q \sla{A}_{\rm osc}  \sla{k} }{2 \pik } \right] 			\gamma^\mu 
								  			\left[ \mathbb{I}_4 + \frac{q \sla{k} \sla{A}_{\rm osc} }{2 \pik} \right] 
						\right\rbrace \varphi . 
\label{eq:term_6}
\end{align}
Substituting \Eqs{eq:term_1}-\eq{eq:term_6} into \Eq{eq:lagr_trans} leads to
\begin{multline}
\mcc{L} =  -\frac{i}{2}\left[ \bar{\varphi} 	\sla{k} 	(\pd_\Theta \varphi) - (\pd_\Theta \bar{\varphi}) \sla{k} \varphi	\right]	\\
			+	\bar{\varphi} \left[ \sla{\pi} + \sla{k} \frac{q^2 \avg{ \msf{A}_{\rm osc}^2 }  }{ 2 \pik} - m \mathbb{I}_4 \right]  \varphi +\mc{F}+\mc{G},
\label{eq:lagr:transition}
\end{multline}
where
\begin{equation}
\mc{F} \doteq \frac{i\epsilon }{2}\left[ \bar{\varphi} \gamma^0 \Xi^\dag \gamma^0	\gamma^\mu 		\Xi (d_\mu \varphi) - \cc	\right]	
\end{equation}
and
\begin{widetext}
\begin{multline}
\mc{G} \doteq  
			\bar{\varphi}	\left[
							  \frac{q\chi }{ \msf{\pi} \cdot \msf{k} } ( \sla{\pi}-m \mathbb{I}_4)
							- \frac{q^2 \chi}{\msf{\pi} \cdot \msf{k} } \sla{A}_{\rm osc}	
							-	 \frac{q^3 \chi }{ 2 \pik^2 }  
										\sla{A}_{\rm osc}  \sla{k}\sla{A}_{\rm osc} 
			\right] \varphi 
			-\epsilon \bar{\varphi} 
				\left[ \mathbb{I}_4+ \frac{q \sla{A}_{\rm osc}  \sla{k} }{2 \pik } \right] 	
				(\sla{d} \tilde{\theta} )
				\left[ \mathbb{I}_4+ \frac{q \sla{k} \sla{A}_{\rm osc} }{2 \pik} \right] \varphi 
				\\
			+\frac{i\epsilon }{2} \bar{\varphi} \left\lbrace
				\left[ \mathbb{I}_4+ \frac{q \sla{A}_{\rm osc}  \sla{k} }{2 \pik } \right] 
				\sla{d} \left[ \frac{q \sla{k} \sla{A}_{\rm osc} }{2 \pik} \right]	
				-		d_\mu 	\left[ \frac{q \sla{A}_{\rm osc}  \sla{k} }{2 \pik } \right] 
										\gamma^\mu 
						\left[ \mathbb{I}_4 + \frac{q \sla{k} \sla{A}_{\rm osc} }{2 \pik} \right] 
				\right\rbrace \varphi . 
\end{multline}
\end{widetext}
Here we introduced $\chi(\epsilon \msf{x}, \Theta) \doteq \msf{k}(\epsilon \msf{x}) \cdot \msf{A}_{\rm osc}(\epsilon \msf{x}, \Theta)$. From \Eqs{eq:lorentz} and \eq{eq:notation_der}, one has $k_\mu \pd_\Theta A_{\rm osc}^\mu = \epsilon d_\mu A^\mu_{\rm osc}$, so
\begin{equation}
	\chi = \epsilon  \int^\Theta d_\mu A^\mu_{\rm osc}(\epsilon \msf{x}, \Theta')\,  \mathrm{d} \Theta'.
\label{eq:dephase}
\end{equation}
Hence, it is seen that $\chi = \mc{O}(\epsilon)$, so $\mc{G} = \mc{O}(\epsilon)$.

\subsection{Approximate Lagrangian density}

The reduced Lagrangian density $\mc{L}$ that governs the time-averaged, or ponderomotive, dynamics can be derived as the time average of $\mcc{L}$, as usual \cite{Whitham:1965kx,Whitham:2011kb}. In our case, the time average coincides with the $\Theta$-average, so
\begin{equation}
\mc{L} \doteq \langle \mcc{L} \rangle.
\end{equation}

Remember that we are interested in calculating $\mc{L}$ with accuracy up to $\mc{O}(\epsilon)$. Using \Eqs{eq:KK} and \eq{eq:asymptotic} and also the fact that $\chi$ is shifted in phase from $\msf{A}_{\rm osc}$ by $\pi/2$ [cf. \Eq{eq:dephase}], it can be shown that $\langle\mc{G}\rangle = \mc{O}(\epsilon^2)$. Therefore, the contribution of $\mc{G}$ to $\mc{L}$ can be neglected. Similarly, we can also neglect the first term in \Eq{eq:lagr:transition} since
\begin{multline}
	-\frac{i}{2}\avg{ \bar{\varphi} 	\sla{k} 	(\pd_\Theta \varphi) 
			- (\pd_\Theta \bar{\varphi}) \sla{k} \varphi	 }		\\
					 =\sum_{n=-\infty }^\infty n \epsilon^{|2n|} \bar{\phi}_n \sla{k} \phi_n = \mc{O}(\epsilon^2),
\end{multline}
where we substituted the asymptotic expansion \eq{eq:asymptotic}. The second term in \Eq{eq:lagr:transition} gives
\begin{multline}
\avg{ \bar{\varphi} \left[ \sla{\pi} + \sla{k} \frac{q^2 \avg{ \msf{A}_{\rm osc}^2 }  }{ 2 \pik} - m \mathbb{I}_4 \right]  \varphi } \\ 
		= \bar{\varphi}_0 \left[ \sla{\pi} + \sla{k} \frac{q^2 \avg{ \msf{A}_{\rm osc}^2 }  }{ 2 \pik} - m \mathbb{I}_4 \right]  \varphi_0 + \mc{O}(\epsilon^2).
\end{multline}
By following similar considerations, we also calculate $\avg{\mc{F}}$, namely as follows. Averaging the first term in $\mc{F}$, we obtain
\begin{align}
&\left\langle  \bar{\varphi}_0  \gamma^0 \Xi^\dag \gamma^0  
		\gamma^\mu \Xi d_\mu \varphi_0  \right\rangle 	\notag	\\
	&	~~	=		\bar{\varphi}_0 \avg{ \left[ \mathbb{I}_4+ \frac{q \sla{A}_{\rm osc}  \sla{k} }{2 \pik } \right]
								\gamma^\mu 
							\left[ \mathbb{I}_4+ \frac{q  \sla{k}  \sla{A}_{\rm osc} }{2 \pik } \right] } d_\mu \varphi_0
							\notag	\\
	&	~~	=		\bar{\varphi}_0 \left[ \gamma^\mu 
										+ \frac{q^2 }{4 \pik^2 }  \avg{ \sla{A}_{\rm osc}  
													\sla{k}  \gamma^\mu \gamma^\nu k_\nu  \sla{A}_{\rm osc} } \right]
														d_\mu \varphi_0
							\notag	\\
	&	~~	=		\bar{\varphi}_0 \left[ \gamma^\mu 
										+ \frac{q^2 }{4 \pik^2 }    \avg{ \sla{A}_{\rm osc}  \sla{k} 
										\left( 2 k^\mu   - \sla{k} \gamma^\mu \right)  \sla{A}_{\rm osc} }	\right]
														d_\mu \varphi_0
							\notag	\\
	&	~~	=		\bar{\varphi}_0 \left[ \gamma^\mu 
										+ k^\mu \frac{q^2 }{2 \pik^2 }   
										 \avg{ \sla{A}_{\rm osc}  \sla{k}   \sla{A}_{\rm osc} }	\right]	d_\mu \varphi_0
							\notag	\\
	&  ~~	=		\bar{\varphi}_0 \Gamma^\mu 	d_\mu \varphi_0,
\end{align}
where we used \Eqs{eq:anticommutation} and \eq{eq:KK}. We also introduced the modified Dirac matrices
\begin{align}
\Gamma^\mu(\epsilon \msf{x} ) 	
		\doteq & \gamma^\mu + k^\mu \frac{q^2}{2  \pik^2 }  
					\avg{  	 \slashed{A}_{\rm osc} \slashed{k} \slashed{A}_{\rm osc} } \notag \\
		= & \gamma^\mu + k^\mu \frac{q^2}{2  \pik^2 }  
					\avg{  	 \slashed{A}_{\rm osc}( 2 \chi- \sla{A}_{\rm osc} \sla{k} )  } \notag \\		
		= & \gamma^\mu - k^\mu \frac{q^2}{2  \pik^2 }  
					\avg{  	 \slashed{A}_{\rm osc} \sla{A}_{\rm osc}  } \sla{k}   \notag \\		
		= & \gamma^\mu - k^\mu \frac{q^2 \avg{ \msf{A}_{\rm osc}^2 }}{2  \pik^2 }   \sla{k}.
\label{eq:Gamma}
\end{align}
Gathering the previous results, we obtain the following reduced Lagrangian density
\begin{multline}
\mc{L}  =  \frac{i}{2}  \left[  \bar{\phi}  \Gamma^\mu  ( \pd_\mu \phi)   - ( \pd_\mu  \bar{\phi} ) \Gamma^\mu \phi   \right] \\
 					+ \bar{\phi} \left[ \sla{\pi} 
 											+ \sla{k} \frac{q^2 \avg{ \msf{A}_{\rm osc}^2 }  }{ 2 \pik} 
 											- m \mathbb{I}_4 
 						\right] \phi
 					+ O(\epsilon^2) ,
  \label{eq:lagr_av_II}
\end{multline}
where $\phi\doteq \varphi_0$. Since only slow spacetime dependences appear in \Eq{eq:lagr_av_II}, we dropped the ``$\epsilon d_\mu$" notation for slow spacetime derivatives and returned to the ``$\pd_\mu$" notation.

\section{Reduced Model}
\label{sec:reduced}

In this section, the Lagrangian density \eq{eq:lagr_av_II} is further simplified by considering only positive-kinetic-energy particle states. The resulting model describes two-component wave functions instead of four-component wave functions, which leads to explicit identification of the spin-coupling term.

\subsection{Particle and antiparticle states}
\label{sec:eigen}

First let us briefly review the case when $\epsilon$ is vanishingly small so that $\pd_\mu\phi$ can be neglected. In this case, \Eq{eq:lagr_av_II} can be approximated as
\begin{equation}
\mc{L}_0 [ \theta, \phi, \bar{\phi}]  = \bar{\phi} \left[ \sla{\pi} + \sla{k} \frac{q^2 \avg{ \msf{A}_{\rm osc}^2}  }{ 2 \pik} - m \mathbb{I}_4 \right]  \phi.
\label{eq:lagr_zero}
\end{equation}
where  $\phi$, $\bar{\phi}$, and $\theta$ can be treated as independent variables. [The Lagrangian density $\mc{L}_0$ depends on $\theta$ in the sense that it depends on $\pi_\mu$, which is defined through $\pd_\mu \theta$ (\Sec{sec:parameterization}).] When varying the action with respect to $\bar{\phi}$, the corresponding ELE is
\begin{equation}\label{eq:eigen_II}
	\delta \bar{\phi}:	\quad	\left( \sla{\lambda} - m \mathbb{I}_4 \right) \phi=0,
\end{equation}
where 
\begin{equation}\label{eq:lambda}
\msf{\lambda}^\mu(\epsilon \msf{x}) \doteq \pi^\mu   + \alpha k^\mu
\end{equation}
is a quasi four-momentum \cite{Yakovlev:1966tp} and
\begin{equation} \label{eq:alpha}
\alpha(\epsilon \msf{x}) \doteq \frac{q^2 \avg{ \msf{A}_{\rm osc}^2}  }{ 2 \pik}.
\end{equation}
The local eigenvalues are obtained by solving
\begin{equation}\label{eq:dispersion}
\det \left(  \sla{\lambda} - m \mathbb{I}_4 \right)=0.
\end{equation}
Since the local dispersion relation \eq{eq:dispersion} has the same form as that of the free-streaming Dirac particle \cite{Thaller:1992th}, one has
\begin{equation}
\lambda \cdot \lambda = \pi\cdot \pi + q^2 \avg{  \msf{A}_{\rm osc}^2} = m^2 ,
\label{eq:dispersion_2}
\end{equation}
where we used \Eq{eq:k_dispersion}. Solving for $\pi^0$ leads to
\begin{equation}\label{eq:explicit}
\pi^0 = -\pd_t \theta - qV_{\rm bg}= \pm \sqrt{  (\del \theta- q\vec{A}_{\rm bg} )^2 +m_{\rm eff}^2 } .
\end{equation}
Here $m_{\rm eff}$ is the ``effective mass'' \cite{Akhiezer:1956wb,Kibble:1966fj, Kibble:1966eb} given by
\begin{equation} \label{eq:mass}
	m^2_{\rm eff} (\epsilon \msf{x}) \doteq m^2- q^2 \avg{ \msf{A}_{\rm osc}^2 }(\epsilon \msf{x}).
\end{equation}
(The minus sign is due to the chosen metric signature.) Equation \eq{eq:explicit} is the well known Hamilton-Jacobi equation that governs the ponderomotive dynamics of a relativistic spinless particle interacting with an oscillating EM vacuum field and a slowly varying background EM field \cite{Malka:1997tc,Quesnel:1998tj,Mora:1997ku,Dodin:2003hz}. The two roots in \Eq{eq:explicit} represent solutions for the particle and the antiparticle states.

\subsection{Eigenmode decomposition}
\label{sec:eigen_decomposition}

Corresponding to the eigenvalues given by \Eq{eq:explicit}, there exists four orthonormal eigenvectors $h_q$ which are obtained from \Eq{eq:eigen_II} and represent the particle and the antiparticle states. Since $h_q$ form a complete basis, one can write $\phi = h_q \eta^q$, where $\eta^q$ are scalar functions. Recall also that pair production is neglected in our model due to the assumption \eq{eq:ordering1}. Let us hence focus on particle states, merely for clarity, which correspond to positive kinetic energies
\begin{equation} \label{eq:Eeff}
\varepsilon_{\rm eff} =  \sqrt{\vec{\pi}^2 + m^2_{\rm eff} }
\end{equation}
in the limit of vanishing $\epsilon$. We will assume that only such states are actually excited (we call these eigenmodes ``active''), whereas the antiparticle states acquire nonzero amplitudes only through the medium inhomogeneities (we call these eigenmodes ``passive''). When designating the active mode eigenvectors by $h_{1,2}$ and the passive mode eigenvectors by $h_{3,4}$, we have
\begin{equation}
\eta^q = \left\{
\begin{array}{ll}
\mc{O}(\epsilon^0), & q=1,2 \\
\mc{O}(\epsilon^1), & q=3,4.\\
\end{array}
\right.
\end{equation}
As shown in \Ref{Ruiz:2015hq}, due to the mutual orthogonality of all $h_q$, the contribution of passive modes to $\mc{L}$ is $o(\epsilon)$, so it can be neglected entirely. In other words, for the purpose of calculating $\mc{L}$, it is sufficient to adopt $\phi \approx h_1 \eta^1 + h_2 \eta^2$. It is convenient to write this active eigenmode decomposition in a matrix form
\begin{equation}\label{eq:proj}
\phi(\epsilon \msf{x}) = \Psi  \eta,
\end{equation}
where
\begin{equation} \label{eq:projector}
\Psi(\epsilon \msf{x})=\sqrt{\frac{m+\lambda^0 }{2\varepsilon_{\rm eff} } } 
\begin{pmatrix}
\mathbb{I}_2 \\
\frac{\vec{\sigma} \cdot \vec{\lambda}}{m+\lambda^0 }
\end{pmatrix}
\end{equation}
is a $4\times 2$ matrix having $h_{1}$ and $h_{2}$ as its columns and
\begin{equation}
\eta(\epsilon \msf{x})\doteq 
\begin{pmatrix}
	\eta^1 \\ 
	\eta^2
\end{pmatrix}.
\end{equation}
Note that $\eta^1(\epsilon \msf{x})$ and $\eta^2(\epsilon \msf{x})$ describe wave envelopes corresponding to the spin-up and spin-down states. 

When inserting the eigenmode representation \eq{eq:proj} into \Eq{eq:lagr_av_II}, one obtains \cite{Ruiz:2015hq}
\begin{equation}\label{eq:lagr_proj}
\mc{L}  = \mc{K} - \eta^\dag \left( \mc{E} - \mc{U} \right) \eta +o(\epsilon),
\end{equation}
where
\begin{gather}
\mc{K} \doteq \frac{i}{2} \left[  \eta^\dag \Psi^\dag \gamma^0  \Gamma^\mu  \Psi ( \pd_\mu \eta)   - \cc  \right] , 		\label{eq:kinetic}	\\
\mc{E} \doteq  \pd_t \theta+ \varepsilon_{\rm eff} +qV_{\rm bg} ,			\label{eq:energy}		\\
\mc{U} \doteq \frac{i}{2} \left[ \Psi^\dag \gamma^0  \Gamma^\mu  ( \pd_\mu \Psi  )   - \hc  \right].		\label{eq:potential}
\end{gather}
The terms $\mc{K}$ and $\mc{U}$, which are of order $\epsilon$, represent corrections to the lowest-order (in $\epsilon$) Lagrangian density. Specifically, for $\mc{K}$ one obtains (\App{app:reduced_kinetic})
\begin{equation}
\mc{K}= \frac{i}{2} \left[ \eta^\dag ( \mathrm{d}_t \eta) - ( \mathrm{d}_t \eta^\dag) \eta \right],
\label{eq:convection}
\end{equation}
where $ \mathrm{d}_t \doteq \pd_t + \vec{v}_0 \cdot \del$ is a convective derivative associated to the zeroth-order velocity field
\begin{equation}
\vec{v}_0(\epsilon \msf{x}) \doteq  \frac{\pd \varepsilon_{\rm eff}}{\pd \vec{p}} = \frac{\vec{\pi}}{\varepsilon_{\rm eff}}.
\label{eq:velocity_zero}
\end{equation}
Regarding $\mc{U}$, one obtains the  ponderomotive spin-orbit coupling Hamiltonian (\App{app:reduced_potential})
\begin{equation}
\mc{U}= \frac{1}{2} \vec{\sigma} \cdot \vec{\Omega}_{\rm eff},
\label{eq:potential_II}
\end{equation}
where
\begin{widetext}
\begin{multline}
\vec{\Omega}_{\rm eff} ( \epsilon \msf{x}) = 
						 \frac{q}{\varepsilon_{\rm eff}} 	
						 	\left( \vec{B}_{\rm bg}   - \frac{\vec{\lambda} \times \vec{E}_{\rm bg} }{m+\lambda^0 }  
						\right)
			 + \frac{q^2 }{2 \varepsilon_{\rm eff} \pik } \left[
						\vec{k}\times  \del \avg{ \msf{A}_{\rm osc}^2} 
						- \frac{(\vec{\lambda} \times \vec{k} ) \pd_t \avg{ \msf{A}_{\rm osc}^2 } }{m+\lambda^0}
						- \frac{ k^0   \vec{\lambda} \times \del \avg{ \msf{A}_{\rm osc}^2}  }{m+\lambda^0} 
						\right]  \\
			 + \frac{q^2 \avg{  \msf{A}_{\rm osc}^2 } }{ 2 \varepsilon_{\rm eff} \pik^2}  
						\left\lbrace		\left( \frac{k^0 \vec{\lambda}}{ m+\lambda^0}  -\vec{k} \right)  \times 
						\left[  k^0  q\vec{E}_{\rm bg} + \vec{k} \times  q\vec{B}_{\rm bg}  
										-(\msf{\pi}^\mu	\pd_\mu)\vec{k}		\right]
				- \frac{ \vec{\lambda}	\times \vec{k} }{m+\lambda^0} 	
						\left[	\vec{k}\cdot q \vec{E}_{\rm bg} - (\msf{\pi}^\mu \pd_\mu) k^0	\right]
						\right\rbrace
\label{eq:omega_eff}
\end{multline}
\end{widetext}
and $\pi \cdot \msf{k} = \varepsilon_{\rm eff} \left( \omega - \vec{k} \cdot \vec{v}_0 \right)$.

When substituting \Eqs{eq:Eeff}, \eq{eq:convection}, and \eq{eq:potential_II} into \Eq{eq:lagr_proj}, one obtains the following effective Lagrangian density
\begin{multline}\label{eq:lagr_proj_II}
\mc{L}  = -\eta^\dag \left( \pd_t \theta + \sqrt{\vec{\pi}^2 + m^2_{\rm eff} } + qV_{\rm bg} \right) \eta \\
						+\frac{i}{2} \left[ \eta^\dag (\mathrm{d}_t \eta) - (\mathrm{d}_t \eta^\dag) \eta \right]
						+ \frac{1}{2} \eta^\dag \vec{\sigma}\cdot \vec{\Omega}_{\rm eff} \eta.
\end{multline}
The first line of \Eq{eq:lagr_proj_II} represents the zeroth-order Lagrangian density that would describe a spinless relativistic electron. The second line, which is of order $\epsilon$, introduces spin-orbit coupling effects. Also note that the Lagrangian density \eq{eq:lagr_proj_II} is analogous to that describing circularly-polarized EM waves in isotropic dielectric media when polarization effects are included \cite{Ruiz:2015dv}.

\section{Continuous wave model}
\label{sec:wave}

Here we construct a ``fluid" description of the Dirac electron described by \Eq{eq:lagr_proj_II}. Let us adopt the representation $\eta = z\sqrt{\mc{I}}$, where $\mc{I}( \msf{x} ) \doteq \eta^\dag \eta$ is a real function (called the action density) and $z( \msf{x} )$ is a unit vector such that $z^\dag z \equiv 1$. [From now on, we drop $\epsilon$ in the function arguments to simplify the notation, but we will continue to assume that the corresponding functions are slow.] Since the common phase of the two components of $z$ can be attributed to $\theta$, we parameterize $z$ in terms of just two real functions $\zeta(\msf{x})$ and $\vartheta(\msf{x})$:
\begin{gather}
z(\vartheta, \zeta) = 
	\begin{pmatrix}
			e^{-i\vartheta/2} \cos (\zeta/2) \\
			e^{i\vartheta/2} \sin (\zeta/2)
	\end{pmatrix}.
\end{gather}
As in the case of the Pauli particle \cite{Ruiz:2015bz}, $\zeta$ determines the relative fraction of ``spin-up'' and ``spin-down'' quanta. Note that, under this reparameterization, the spin vector $\vec{S}(\msf{x})$ is given by
\begin{equation}
\vec{S}	\doteq	\frac{1}{2} z^\dag \vec{\sigma} z =	
		\begin{pmatrix}
				\sin\zeta \cos\vartheta 	\\
				\sin\zeta \sin\vartheta		\\
				\cos\zeta
		\end{pmatrix},
\end{equation}
and $S\doteq |\vec{S}|=1/2$. 

Expressing \Eq{eq:lagr_proj_II} in terms of the four independent variables $(\theta, \mc{I}, \zeta, \vartheta)$ leads to
\begin{multline}\label{eq:lagr_wave}
\mc{L}[\theta,\mc{I}, \zeta,\vartheta]  = -\mc{I} \left[ \pd_t \theta + \sqrt{\vec{\pi}^2 + m^2_{\rm eff} } + qV_{\rm bg} \right. \\ 
					\left.
							-\frac{1}{2} (\mathrm{d}_t \vartheta)	\cos \zeta - \vec{S}(\zeta, \vartheta) \cdot \vec{\Omega}_{\rm eff}
					\right],
\end{multline}
where one can immediately recognize the first line of \Eq{eq:lagr_wave} as Hayes' representation of the Lagrangian density of a GO wave \cite{Hayes:1973dt}. Four ELEs are yielded. The first one is the action conservation theorem
\begin{gather}\label{eq:act}
\delta \theta : \quad \pd_t \mc{I} + \del \cdot ( \mc{I} \vec{V}) = 0.
\end{gather}
The flow velocity is given by $\vec{V} = \vec{v}_0 + \vec{u}$, where
\begin{gather}
\vec{u} \doteq -
		\frac{\pd}{\pd\vec{p}}	\left[	\frac{1}{2}	(\vec{v}_0 \cdot \del \vartheta) \cos\zeta  
																	+ \vec{S}(\zeta, \vartheta) \cdot \vec{\Omega}_{\rm eff} \right]  
\end{gather}
is the spin-driven deflection of the electron's center of mass. The second ELE is a Hamilton-Jacobi equation
\begin{multline}
\delta \mc{I} : \quad   \pd_t \theta + \sqrt{\vec{\pi}^2 + m^2_{\rm eff} } + qV_{\rm bg}  \\
							-\frac{1}{2} (\mathrm{d}_t \vartheta)	\cos \zeta - \vec{S}(\zeta, \vartheta) \cdot \vec{\Omega}_{\rm eff}	=0 ,
\end{multline}
whose gradient yields the momentum equation
\begin{multline}
\pd_t \vec{\pi}+(\vec{v}_0 \cdot \del) \vec{\pi} = q\vec{E}_{\rm bg} + q \vec{v}_0 \times \vec{B}_{\rm bg} \\
	+\frac{\del \avg{\msf{A}_{\rm osc}^2 } }{2 \varepsilon_{\rm eff}  } 
	+\del \left[ 	\frac{1}{2} (\mathrm{d}_t \vartheta)	\cos \zeta+ \vec{S} \cdot \vec{\Omega}_{\rm eff}	\right].
	\label{eq:momentum}
\end{multline}
Note that the first line is the well known relativistic momentum equation. The first term in the second line represents the well known nonlinear ponderomotive force due to the oscillating EM field \cite{Mora:1997ku} and the last two terms represent the ponderomotive Stern-Gerlach spin force. Finally, the remaining two ELEs are
\begin{align}
\delta \zeta :	& \quad 
			(	\mathrm{d}_t \vartheta )	\sin \zeta	
						=2 (\pd_\zeta \vec{S}) \cdot \vec{\Omega}_{\rm eff},  \label{eq:zeta}		\\
\delta \vartheta : &	\quad 		
			\pd_t( \mc{I} \cos \zeta) +  \del \cdot (\vec{v}_0	\mc{I} \cos \zeta ) 
						=2 (\pd_\vartheta \vec{S}) \cdot \vec{\Omega}_{\rm eff}. \label{eq:vartheta}
\end{align}
These equations describe the phase-averaged electron spin precession. Together, \Eqs{eq:act}-\eq{eq:vartheta} provide a complete ``fluid'' description of the ponderomotive dynamics of a Dirac electron.

\section{Point-particle model}
\label{sec:point}

\subsection{Ponderomotive model}
\label{sec:point_pond}

The ray equations corresponding to the above field equations can be obtained as a point-particle limit. In this limit, $\mc{I}$ can be approximated with a delta function,
\begin{gather}
\mc{I}(t, \vec{x}) = \delta(\vec{x} - \vec{X}(t)),
\end{gather}
where $\vec{X}(t)$ is the location of the center of the wave packet. As in \Refs{Ruiz:2015bz,Ruiz:2015hq}, the Lagrangian density \eq{eq:lagr_wave} can be replaced by a point-particle Lagrangian $L_{\rm eff}	\doteq \int \mc{L}~ \mathrm{d}^3x$, namely,
\begin{multline}
L_{\rm eff}[ \vec{X},\vec{P},Z,Z^\dag] = \vec{P} \cdot \dot{\vec{X}}  
			+\frac{i \hbar}{2}  \left( Z^\dag \dot{Z} - \dot{Z}^\dag Z \right) \\
			-H_{\rm eff}( t,\vec{X},\vec{P},Z,Z^\dag) ,
\label{eq:lagr_point_pond}
\end{multline}
where the effective Hamiltonian is given by
\begin{equation}
H_{\rm eff}( t, \vec{X},\vec{P},Z,Z^\dag) \doteq  \gamma_{\rm eff}    mc^2 + q\vec{V}_{\rm bg} 
					-\frac{\hbar}{2} Z^\dag \vec{\sigma}\cdot \vec{\Omega}_{\rm eff} Z.
\label{eq:H_point_pond}
\end{equation}
Here $\vec{P}( t ) \doteq \hbar \del \theta( t , \vec{X}(t) )$ is the canonical momentum, and $Z(t)\doteq z(t,\vec{X}(t))$ is a complex two-component vector. For clarity, we have re-introduced $c$ and $\hbar$. The effective Lorentz factor associated with the particle oscillation-center motion is
\begin{equation}
\gamma_{\rm eff}(t,\vec{X},\vec{P})	\doteq   \sqrt{ 1	+	a_0^2
			+ \left( \frac{\vec{P}}{mc} -\frac{q\vec{A}_{\rm bg}}{mc^2} \right)^2  } ,
\label{eq:lorentz_pond}
\end{equation}
where
\begin{equation}
a_0^2(t,\vec{X}) \doteq -\frac{q^2 \avg{ \msf{A}_{\rm osc}^2 } }{ m^2c^4}
\label{eq:a_0}
\end{equation}
is positive under the assumed metric. For example, suppose a standard representation of the laser vector potential is $\vec{A}_{\rm osc} = \mathrm{Re} \left[ \vec{A}_\perp (\msf{x}) e^{i\Theta} \right]$, where $\vec{A}_\perp \cdot \vec{k}=0$ \cite{Mora:1996jy,Malka:1997tc,Mora:1998zz}. Then, the Lorentz condition \eq{eq:lorentz} determines the  scalar potential envelope $V_{\rm osc,c} = i (\del \cdot \vec{A}_\perp ) c^2 / \omega=\mc{O}(\epsilon)$. Hence, \Eq{eq:a_0} yields

\begin{gather}
	a_0^2 \approx \frac{q^2|\vec{A}_\perp|^2}{2m^2c^4},
\end{gather}
where we neglected a term $\mc{O}(\epsilon^2)$. Note also that, loosely speaking, $a_0^2$ is the measure of the particle quiver energy in units $mc^2$. Accordingly, nonrelativistic interactions correspond to $a_0 \ll 1$.

The effective precession frequency $\vec{\Omega}_{\rm eff}$ is given by
\begin{equation}
\vec{\Omega}_{\rm eff} (t,\vec{X},\vec{P}) = 
		\vec{\Omega}_1
	+ 	\vec{\Omega}_2
	+	\vec{\Omega}_3
	+ \mc{O}(\epsilon^2),
\end{equation}
where
\begin{equation}
\vec{\Omega}_1 (t,\vec{X},\vec{P}) \doteq 
		 \frac{q}{\gamma_{\rm eff} mc} \left(  \vec{B}_{\rm bg}  
						 			- \frac{\vec{\Lambda} \times \vec{E}_{\rm bg} }{mc+\Lambda^0}   \right),
\end{equation}
\begin{multline}
\vec{\Omega}_2 (t,\vec{X},\vec{P}) \doteq 
			- \frac{m c^2 }{2  \gamma_{\rm eff} \Pik } \bigg[
						\vec{k}\times  \del a_0^2 \\
						- \frac{(\vec{\Lambda} \times \vec{k} ) \pd_t  a_0^2 }{mc^2+\Lambda^0c} 
						- \frac{ \omega   \vec{\Lambda} \times \del a_0^2  }{mc^2+\Lambda^0 c}
						\bigg],	
\end{multline}
\begin{multline}
\vec{\Omega}_3 (t,\vec{X},\vec{P}) \doteq 
			- \frac{ m c^2 a_0^2  }{2 \gamma_{\rm eff}  \Pik^2} 
						\left( \frac{\omega \vec{\Lambda}}{ mc^2+\Lambda^0 c}  -\vec{k} \right)  \\
						\times 
						\left[  \frac{ \omega  q\vec{E}_{\rm bg} }{c^2} 
								+ \frac{\vec{k} \times  q\vec{B}_{\rm bg} }{c}
								- (\Pi^\mu \pd_\mu) \vec{k} 	\right]		 \\
					+	\frac{ m c ~a_0^2 }{ 2 \gamma_{\rm eff} \Pik^2}  
			  				\frac{\vec{\Pi}\times \vec{k} }{mc+\Lambda^0} 
			  				\left[	\vec{k}\cdot q\vec{E}_{\rm bg} 	- (\Pi^\mu \pd_\mu) \omega \right].
\end{multline}
Here $\Pi^\mu=(mc\gamma_{\rm eff}, \vec{P}-q\vec{A}_{\rm bg}/c )$, $k^\mu=(\omega/c,\vec{k})$, $\pd_\mu = ( c^{-1}\pd_t, \del)$, and
\begin{equation}
\Lambda^\mu (t,\vec{X},\vec{P}) =\Pi^\mu - k^\mu \frac{m^2 c^2 a_0^2}{2 \Pik } .
\label{eq:Lambda_pond}
\end{equation}
Notably, $\Lambda^\mu \to \Pi^\mu$ at $a_0 \to 0$ and $\Lambda^\mu \to \Pi^\mu - k^\mu mc^2 a_0/(2\omega)$ at $a_0	\to +\infty$. Also, if the spin-orbital interaction is neglected, the present model yields the spinless ponderomotive model that was developed in \Ref{Dodin:2003hz} for a particle interacting with a laser pulse and a slow background fields simultaneously.

\begin{figure*}
  	\includegraphics[width=\textwidth]{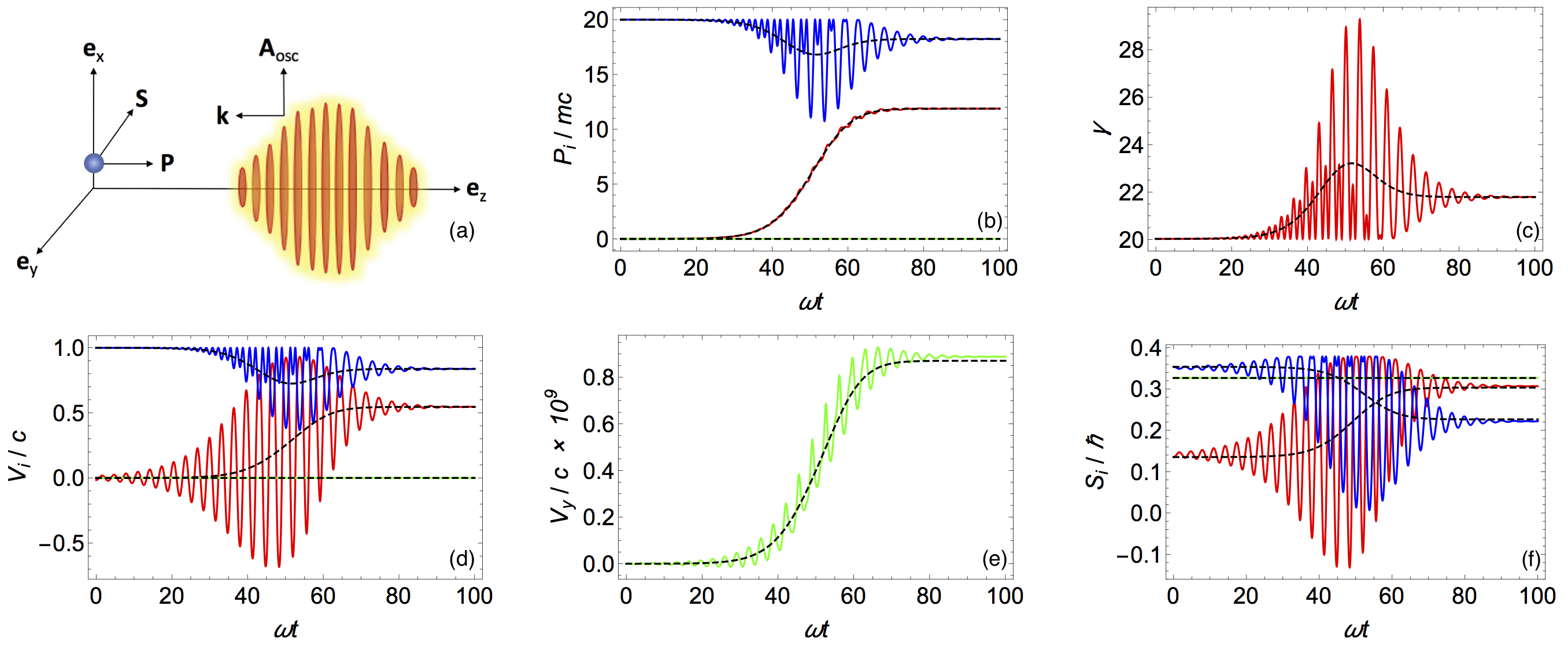}
  	\caption{
Motion of a single Dirac electron under the action of a relativistically intense laser pulse (numerical simulation): the black dashed curves correspond to the ponderomotive model described by the Lagrangian \eq{eq:lagr_point_pond}, and the colored curves correspond to the XBMT model described by the Lagrangian \eq{eq:lagr_point}. (a) Schematic of the interaction; yellow and red is the laser field, blue is the particle; arrows denote the direction of the laser wave vector $\vec{k}$, the oscillating vector potential $\vec{A}_{\rm osc}$, the particle canonical momentum $\vec{P}$, and the particle spin $\vec{S}$. The unit vectors along the reference axes are denoted by $\vec{e}_i$. Figures (b)-(f) show the components  of the particle canonical momentum $\vec{P}$, Lorentz factor $\gamma$, velocity $\vec{V}$, and spin $\vec{S}$. The red, green, and blue lines correspond to projections on the $x$, $y$, and $z$ axes, respectively. We consider an electron initially traveling along the $z$-axis and colliding with a counter-propagating laser pulse. 
The initial position of the particle is $\vec{X}_0=(\ell/2) \vec{e}_x$, the initial momentum is $\vec{P}_0/(mc)=20 \vec{e}_z$, and the normalized initial spin vector is $\vec{S}_0/\hbar=0.14\vec{e}_x +0.33\vec{e}_y +0.35\vec{e}_z$. The envelope of the vector potential of the laser pulse is $q\vec{A}_{\rm osc}/(mc^2) = 30 ~\mathrm{sech} \left[ (z-5\ell +ct)/\ell \right] \exp\left[ -(x^2+y^2)/\ell^2 \right] \vec{e}_x$, where $\ell=20 |\vec{k}|^{-1}$. These parameters correspond to a maximum intensity $I_{\rm max}\simeq 1.23 \times 10^{21}$ W/cm$^2$ for a $1\,\mu$m laser.
}
  	\label{fig:image_Ia}
\end{figure*}

Treating $\vec{X}(t)$, $\vec{P}(t)$, $Z(t)$, and $Z^\dag(t)$ as independent variables leads to the following ELEs:
\begin{align}
\delta \vec{P} : & \quad \dot{\vec{X}} =  \pd_{\vec{P}} ( \gamma_{\rm eff} mc^2)
									- \pd_{\vec{P}} (\vec{S} \cdot  \vec{\Omega}_{\rm eff} ) , \label{eq:x_pond} \\
\delta \vec{X} : & \quad \dot{\vec{P}} = - \pd_{\vec{X}}  (  \gamma_{\rm eff} mc^2 + qV_{\rm bg} )
									+ \pd_{\vec{X}} (\vec{S} \cdot  \vec{\Omega}_{\rm eff}), \label{eq:p_pond} \\ 
\delta Z^\dag  : & \quad \dot{Z} = \frac{i}{2} \vec{\Omega}_{\rm eff} \cdot \vec{\sigma} Z , 	\label{eq:z_pond}\\
\delta Z       : & \quad \dot{Z}^\dag = -\frac{i}{2} Z^\dag \vec{\Omega}_{\rm eff} \cdot \vec{\sigma} \label{eq:zdag_pond}	,
\end{align}
where $\vec{S}(t)$ is the particle spin vector,
\begin{equation}
\vec{S}(t) \doteq \frac{\hbar}{2} Z^\dag(t) \vec{\sigma} Z(t),
\label{eq:spin_def}
\end{equation}
and $S=\hbar /2$. Equations \eq{eq:lorentz_pond}-\eq{eq:spin_def} form a complete set of equations. The first terms on the right hand side of \Eqs{eq:x_pond} and \eq{eq:p_pond} describe the dynamics of a relativistic spinless particle in agreement with earlier theories \cite{Malka:1997tc,Quesnel:1998tj,Mora:1997ku,Dodin:2003hz}. The second terms describe the ponderomotive spin-orbit coupling. Equations \eq{eq:z_pond} and \eq{eq:zdag_pond} also yield the following ponderomotive equation for spin precession,
\begin{equation}\label{eq:spin}
\dot{\vec{S}}=\vec{S} \times \vec{\Omega}_{\rm eff} ,
\end{equation}
which can be checked by direct substitution. Equations \eq{eq:lagr_point_pond}-\eq{eq:spin} are the main result of this work.

\begin{figure*}
  	\includegraphics[width=\textwidth]{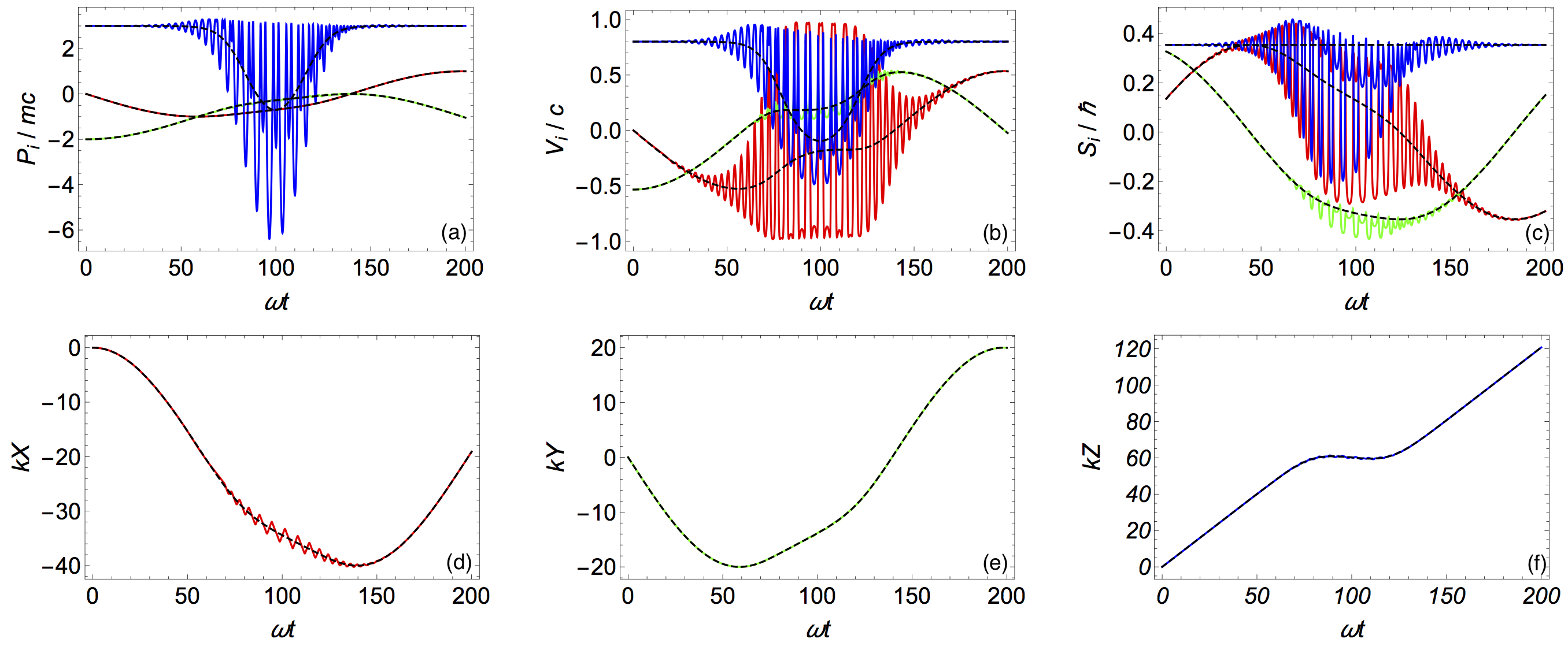}
  	\caption{
Motion of a Dirac electron under the action of an external background field and a relativistically intense laser pulse (numerical simulation): the black dashed curves correspond to the ponderomotive model described by the Lagrangian \eq{eq:lagr_point_pond}, and the colored curves correspond to the XBMT model described by the Lagrangian \eq{eq:lagr_point}. Figures (a)-(c) show the components of the particle canonical momentum $\vec{P}$, velocity $\vec{V}$, and spin $\vec{S}$. The red, green, and blue lines correspond to projections on the $x$, $y$, and $z$ axes, respectively. We consider an electron initially traveling along the $z$-axis and colliding with a counter-propagating laser pulse. Figures (d)-(f) show the components of the particle position $\vec{X}$. 
The initial position of the particle is $\vec{X}_0=\vec{0}$, the initial momentum is $\vec{P}_0/(mc)=(-2\vec{e}_y+3\vec{e}_z)$, and the normalized initial spin vector is $\vec{S}_0/\hbar=0.14\vec{e}_x +0.33\vec{e}_y +0.35\vec{e}_z$.
A background magnetic field is added such that $q\vec{A}_{\rm bg}/(mc^2)=0.1 (-y \vec{e}_x +x\vec{e}_y)/2$, which corresponds to a static homogeneous magnetic field $B_{\rm bg} \simeq 10.7 ~\mathrm{MG}$ aligned towards the $z$-axis.  
The envelope of the vector potential of the laser pulse is assumed to have the form $q\vec{A}_{\rm osc} /(mc^2) =  10~\mathrm{sech} \left[ (z-8\ell +ct)/\ell \right]  \vec{e}_x$, where $\ell=20 |\vec{k}|^{-1}$. 
These parameters correspond to a maximum laser intensity $I_{\rm  max} \simeq 1.37 \times 10^{20}$ W/cm$^2$ for a $1\,\mu$m laser.
}
  	\label{fig:image_II}
\end{figure*}

\subsection{Extended BMT model}

Let us compare our ponderomotive point-particle Lagrangian \eq{eq:lagr_point_pond} with the complete point-particle Lagrangian of a Dirac electron \cite{Ruiz:2015hq}
\begin{multline} 
L_{\rm XBMT}[\vec{X},\vec{P},Z,Z^\dag] =	\vec{P}\cdot \dot{\vec{X}} +\frac{i \hbar }{2}\left( Z^\dag \dot{Z} - \dot{Z}^\dag Z \right) 	\\
	- H_{\rm XBMT}(t,\vec{X},\vec{P},Z,Z^\dag),  
\label{eq:lagr_point}
\end{multline}
where the Hamiltonian is given by
\begin{equation}
H_{\rm XBMT} (t,\vec{X},\vec{P},Z,Z^\dag)\doteq \gamma mc^2 +qV- \frac{\hbar}{2} Z^\dag \vec{\sigma} \cdot \vec{\Omega}_{\rm BMT} Z
\end{equation}
and the BMT precession frequency \cite{Bargmann:1959us} is
\begin{equation}
 \vec{\Omega}_{\rm BMT} (t,\vec{X}, \vec{P}) = \frac{q}{mc} \left[ \frac{\vec{B}}{\gamma }- \frac{ (\vec{v}_0/c) \times \vec{E}  }{1+\gamma}\right],
 \label{eq:omega_BMT}
\end{equation}
Here $\vec{v}_0 \doteq \vec{\Pi}/( {\gamma m}) $, and 
\begin{equation}
\gamma(t,\vec{X},\vec{P}) \doteq \sqrt{ 1+  \left( \frac{\vec{P}}{mc} -\frac{q\vec{A}}{mc^2} \right)^2 }.
\end{equation}
Obviously, $L_{\rm XMBT} \to L_{\rm eff}$ when $a_0 \to 0$.

The corresponding ELEs are
\begin{align}
\delta \vec{P} : & \quad \dot{\vec{X}} =  \pd_{\vec{P}} ( \gamma mc^2 )
										- \pd_{\vec{P}} (\vec{S} \cdot  \vec{\Omega}_{\rm BMT}), \label{eq:x} \\
\delta \vec{X} : & \quad \dot{\vec{P}} = - \pd_{\vec{X}}  (  \gamma mc^2 + qV )
										+\pd_{\vec{X}} ( \vec{S} \cdot  \vec{\Omega}_{\rm BMT} ), \label{eq:p} \\ 
\delta Z^\dag  : & \quad \dot{Z} = \frac{i}{2} \vec{\Omega}_{\rm BMT} \cdot \vec{\sigma} Z , 	\label{eq:z}\\
\delta Z       : & \quad \dot{Z}^\dag = -\frac{i}{2} Z^\dag \vec{\Omega}_{\rm BMT} \cdot \vec{\sigma} \label{eq:zdag}	.
\end{align}
These equations also yield the BMT spin precession equation, similar to \Eq{eq:spin}, with $\vec{\Omega}_{\rm eff}$ replaced by $\vec{\Omega}_{\rm BMT}$. However, as opposed to the original BMT model \cite{Bargmann:1959us,foot:anomalous}, \Eqs{eq:x}-\eq{eq:zdag} also capture the spin-orbital coupling. Because of that, they represent a generalization of the BMT model, which we call ``extended BMT'' (XBMT).

The XBMT model applies, in principle, to arbitrary fields, provided that (i) the spin-orbital coupling is weak and (ii) the particle de Broglie wavelength $\lambdabar$ remains much shorter than the smallest spatial scale of the EM fields. In application to the particle motion in a laser field, it can describe details that the ponderomotive model misses due to phase averaging. In \textit{this} sense, the XBMT model is more precise than the ponderomotive model above. However, the XBMT (and, similarly, BMT) model is also more complicated for the same reason and, in application to laser fields, requires $ \lambdabar / \lambda_L \ll 1$, where $\lambda_L$ is the laser wavelength. No such assumption was made to derive the ponderomotive model above. Instead, \Eq{eq:ordering1} was assumed, which implies
\begin{equation}
\lambdabar/\lambda_L \ll c/v_0,
\end{equation}
where $v_0$ is the particle speed. For nonrelativistic particles ($v_0 \ll c$), this can be satisfied even at $\lambda_L \lesssim \lambdabar$. In \textit{that} sense, the ponderomotive model is, perhaps surprisingly, more general than XBMT.

\section{Numerical simulations}
\label{sec:numerical}

To test our ponderomotive model, we applied it to simulate the single-particle motion and compare the results with the XBMT model in two test cases. In the first test case, we consider the dynamics of a Dirac electron colliding with a counter-propagating relativistically strong ($a_0 \gg 1$) laser pulse. The simulation parameters are given in the caption of Fig. 1, and a schematic of the interaction is presented in Fig. 1(a). From Figs. 1(b)-1(e), it is seen that the ponderomotive model accurately describes the mean evolution of the particle momentum, kinetic energy, and velocity. The main contribution to the variations in $V_x$ and $V_z$ is the ponderomotive force caused by spatial gradient of the effective mass. However, the acceleration on the $xz$-plane is caused by the Stern-Gerlach force, as shown in Fig.~1(e). Also notice that the ponderomotive model is extremely accurate in describing the particle spin precession, as shown in Fig.~1(f). 

In the second test case, we consider a Dirac electron immersed in a background magnetic field along the $z$-axis and interacting with a laser plane wave traveling along the $z$-axis. The simulation parameters are given in Fig.~2. As can be seen in Figs. 2(a)-2(f), the ponderomotive model accurately describes the particle position, momentum, velocity, and spin. Notably, these simulations also support the spinless model developed in \Ref{Dodin:2003hz} for a particle interacting with a relativistic laser field and a large-scale background field simultaneously.

\section{Conclusions}
\label{sec:conclusions}

In this paper, we report a point-particle ponderomotive model of a Dirac electron oscillating in a high-frequency field. Starting from the first-principle Dirac Lagrangian density,  we derived a reduced phase-space Lagrangian that describes the relativistic time-averaged dynamics of such particle in a geometrical-optics laser pulse in vacuum. The pulse is allowed to have an arbitrarily large amplitude (as long as radiation damping and pair production are negligible) and, in case of nonrelativistic interactions, a wavelength comparable to the electron de Broglie wavelength. The model captures the BMT spin dynamics, the Stern-Gerlach spin-orbital coupling, the conventional ponderomotive forces, and the interaction with large-scale background fields (if any). Agreement with the BMT spin precession equation is shown numerically. Also, the well known theory, in which ponderomotive effects are incorporated in the particle effective mass, is reproduced as a special case when the spin-orbital coupling is negligible.

As a final note, the underlying essence of this paper is to illustrate the convenience of using the Lagrangian wave formalism for deriving reduced point-particle models. To derive the ponderomotive model above by using the point-particle equations of motion and spin would have been a torturous task. However, the bilinear structure of the wave Lagrangian enabled a straightforward deduction of the reduced model. Following this reasoning, we believe that the ability to treat particles and waves on the same footing, \ie as fields, may have far-reaching implications, \eg for plasma theory. This will be discussed in future publications.

The authors thank E. A. Startsev for valuable discussions. This work was supported by the NNSA SSAA Program through DOE Research Grant No. DE274-FG52-08NA28553, by the U.S. DOE through Contract No. DE-AC02-09CH11466, by the U.S. DTRA through Research Grant No. HDTRA1-11-1-0037, and by the U.S. DOD NDSEG Fellowship through Contract No. 32-CFR-168a.

\appendix

\section{Semiclassical Volkov state}
\label{app:volkov}

Volkov states are eigenstates of the Dirac equation with an homogeneous EM vacuum field \cite{Volkov:1935vka,Vergou:1979,Raicher:2013ik}. Here we present a derivation of these states. Consider the second order Dirac equation,
\begin{equation}
\left( D_\mu D^\mu +m^2 +\frac{1}{2}q  \sigma_{\mu \nu} F^{\mu \nu} \right) \psi=0,
\label{eq:volkov_order2}
\end{equation}
where $iD_\mu \doteq i\pd_\mu -qA_\mu$ is the covariant derivative, $\sigma_{\mu \nu} \doteq  i[ \gamma_\mu , \gamma_\nu ]/2 $ is twice the (relativistic) spin operator, and $F^{\mu\nu}=\pd^\mu A^\nu - \pd^\nu A^\mu$ is the EM tensor. We start with the case where $\msf{A}_{\rm bg}$ is constant and $\msf{A}_{\rm osc}(\Theta)$ is strictly periodic. Since \Eq{eq:volkov_order2} is linear, we search for $\psi$ in the Floquet-Bloch form. Specifically, we consider $\psi = ue^{i\theta}$, where $u$ is a periodic four-component function of $\Theta$ and $p_\mu \doteq - \pd_\mu \theta$ is constant. It is also convenient to rewrite $u$ in the form $u = e^{i\tilde{\theta}}\Xi \varphi $, where $\Xi(\Theta)$ is a matrix operator, $\tilde{\theta}(\Theta)$ is a real scalar function, and $\varphi$ is a constant four-component spinor. This leads to
\begin{multline}
\left[ \msf{\pi}^2 - m^2 +2 \pik \pd_\Theta \tilde{\theta} -2 q (\msf{\pi} \cdot \msf{A}_{\rm osc}) +q^2 \msf{A}_{\rm osc}^2 \right] \Xi \varphi	\\
	-2i \pik (\pd_\Theta \Xi)	\varphi  -\frac{1}{2}q\sigma_{\mu \nu} F^{\mu \nu} \Xi \varphi =0,
	\label{eq:volkov_semi}
\end{multline}
where $\pi^\mu \doteq p^\mu - q A_{\rm bg}^\mu$. 

Equation \eq{eq:volkov_semi} can be satisfied identically if we require that $\tilde{\theta}$ and $\Xi$ satisfy the following equations:
\begin{gather}
	\msf{\pi}^2 - m^2 +2 \pik \pd_\Theta \tilde{\theta} 
					-2 q (\msf{\pi} \cdot \msf{A}_{\rm osc}) +q^2 \msf{A}_{\rm osc}^2 =0, \label{eq:volkov_first}\\
	-2i \pik (\pd_\Theta \Xi) -\frac{1}{2}q\sigma_{\mu \nu} F^{\mu \nu} \Xi  =0		\label{eq:volkov_second}.
\end{gather}
The integration constants can be chosen arbitrarily since they merely redefine $\varphi$. We hence require $\Xi \to \mathbb{I}_4$ at vanishing $\msf{A}_{\rm osc}$ and $\langle\tilde{\theta}	\rangle = 0$ (so that $\tilde{\theta}$ represents a phase shift due to the oscillating EM field). For $\Xi$, this gives
\begin{align}
\Xi(\epsilon \msf{x}, \Theta) 
			=	&  \mc{T} \exp \left[ \frac{iq}{4 \pik} \int^\Theta  \sigma_{\mu \nu} F^{\mu \nu}(\Theta')  \, d\Theta'  \right]  \notag \\
			=	& 	\mathbb{I}_4+  \frac{q}{2 \pik} \sla{k} \sla{A}_{\rm osc}(\Theta) ,
\label{eq:XI:derivation}
\end{align}
where we used
\begin{align}\label{eq:prop_1}
 \sigma_{\mu \nu} F^{\mu \nu} =  & 
			 \sigma_{\mu \nu} (\pd^\mu A^\nu - \pd^\nu A^\mu )  \notag \\
=& 		-\sigma_{\mu \nu} (k^\mu \pd_\Theta A^\nu_{\rm osc} - k^\nu \pd_\Theta A^\mu_{\rm osc} )  \notag \\
=&		-i ( \slashed{k} \pd_\Theta \slashed{A}_{\rm osc} -  \pd_\Theta \slashed{A}_{\rm osc} \slashed{k}  )  \notag \\
=& 		-2i \slashed{k} \pd_\Theta \slashed{A}_{\rm osc} .
\end{align}
Here $ \slashed{k} \slashed{A}_{\rm osc} + \slashed{A}_{\rm osc}\slashed{k}=0$ since $\mathsf{k} \cdot \mathsf{A}_{\rm osc}=0$ [see \Eq{eq:lorentz}]. We note that the ordered exponential [denoted by $\mathcal{T}\exp(...)$] becomes an ordinary exponential due to
\begin{align} \label{eq:prop_2}
&\sigma_{\mu \nu} F^{\mu \nu}(\Theta_1) \sigma_{\alpha \xi} F^{\alpha \xi}(\Theta_2)  	\notag \\
&~~~~~~~		=-4 \slashed{k} [\pd_\Theta \slashed{A}_{\rm osc}(\Theta_1)] \slashed{k}
							[\pd_\Theta \slashed{A}_{\rm osc}(\Theta_2)]  \notag \\
&~~~~~~~		= 4 \slashed{k} \slashed{k} \pd_\Theta \slashed{A}_{\rm osc}(\Theta_1) 
									\slashed{A}_{\rm osc}(\Theta_2) \notag \\
&~~~~~~~		= 0,
 \end{align}
where we substituted \Eq{eq:KK}. To obtain $\tilde{\theta}$, we average \Eq{eq:volkov_first}. This leads to \Eq{eq:dispersion}, which serves as a dispersion relation for $\pi_\mu$. Subtracting \Eq{eq:dispersion} from \Eq{eq:volkov_first} and solving for $\tilde{\theta}$ leads to \Eq{eq:Ip}. Finally, if one substitutes $\psi=\Xi e^{i\theta + \tilde{\theta}}\varphi$ into the first-order Dirac equation, one finds that constant $\varphi$ indeed satisfies that equation.

The above solution can be extended also to a wave with a slowly inhomogeneous amplitude; \ie when the vector potential has the form $\msf{A}(\epsilon\msf{x}, \Theta)$. This can be done by substituting the ansatz $\psi=\Xi e^{i\theta + \tilde{\theta}}\varphi$ into the Dirac equation with the same $\Xi$ and $\tilde{\theta}$, as before, and \textit{requiring} that $p_\mu$ is slow. This will lead to an equation for $\varphi$ with a perturbation linear in $\epsilon$. Hence, one can construct a solution for $\varphi$ as an asymptotic power series in $\epsilon$. The general form of such series in given by \Eq{eq:asymptotic}, and finding the coefficients $\varphi_n$ explicitly is not needed here.


\section{Auxiliary Formulas}
\label{app:reduced}


\subsection{Kinetic Term $\boldsymbol{\mc{K}}$ }
\label{app:reduced_kinetic}

Let us re-express \Eq{eq:kinetic} as
\begin{equation}
\mc{K}  =  \frac{i}{2}\left[   \eta^\dag \Psi^\dag \gamma^0 \Gamma^0 \Psi (\pd_t \eta) +  \eta^\dag \Psi^\dag \gamma^0 \vec{\Gamma} \cdot \Psi (\del \eta) -\cc \right] .
\label{eq:kinetic2}
\end{equation}
Substituting \Eqs{eq:Gamma}, \eq{eq:lambda}, \eq{eq:alpha}, and \eq{eq:projector} into $\Psi^\dag \gamma^0 \Gamma^0 \Psi$ leads to
\begin{widetext}
\begin{align} 
\Psi^\dag \gamma^0 \Gamma^0 \Psi = &
 		\frac{m+\lambda^0 }{2\varepsilon_{\rm eff} } \begin{pmatrix}  \mathbb{I}_2 & \frac{\vec{\sigma} \cdot \vec{\lambda}}{m+\lambda^0 } \end{pmatrix}
			\left( \gamma^0 \gamma^0  -    \frac{k^0\alpha }{\msf{\pi} \cdot \msf{k} }  \gamma^0 \sla{k} \right)
		\begin{pmatrix}  \mathbb{I}_2 \\ \frac{ \vec{\sigma} \cdot \vec{\lambda} }{m+ \lambda^0 }  \end{pmatrix} \notag \\
 =&	\frac{m+\lambda^0 }{2\varepsilon_{\rm eff} } \begin{pmatrix}  \mathbb{I}_2 & \frac{\vec{\sigma} \cdot \vec{\lambda}}{m+\lambda^0 } \end{pmatrix}
			\left[ \mathbb{I}_4  -   \frac{k^0\alpha }{\msf{\pi} \cdot \msf{k} }
							\begin{pmatrix}  k^0 & - \vec{\sigma} \cdot \vec{k} \\ - \vec{\sigma} \cdot \vec{k}  & k^0 \end{pmatrix}			
			 \right]
		\begin{pmatrix}  \mathbb{I}_2 \\ \frac{ \vec{\sigma} \cdot \vec{\lambda} }{m+ \lambda^0 }  \end{pmatrix} \notag \\
 =&	\frac{m+\lambda^0 }{2\varepsilon_{\rm eff} } \begin{pmatrix}  \mathbb{I}_2 & \frac{\vec{\sigma} \cdot \vec{\lambda}}{m+\lambda^0 } \end{pmatrix}
			\left[ 
			\begin{pmatrix}  \mathbb{I}_2 \\ \frac{ \vec{\sigma} \cdot \vec{\lambda} }{m+ \lambda^0}  \end{pmatrix} 
			 -   \frac{k^0\alpha }{\msf{\pi} \cdot \msf{k} }
							\begin{pmatrix}  k^0 \mathbb{I}_2 -  \frac{ (\vec{\sigma} \cdot \vec{k}) (\vec{\sigma} \cdot \vec{\lambda}) }{m+ \lambda^0}  \\
							  - \vec{\sigma} \cdot \vec{k} + k^0 \frac{ \vec{\sigma} \cdot \vec{\lambda} }{m+ \lambda^0 } \end{pmatrix}			
			 \right] \notag \\
 =&	\frac{m+\lambda^0 }{2\varepsilon_{\rm eff} } 	\left\lbrace
			  1 +  \frac{ \vec{\lambda}^2 }{(m+ \lambda^0)^2} 
			 -   \frac{k^0\alpha }{\msf{\pi} \cdot \msf{k} }
							\left[  k^0 -  \frac{ 2\vec{k} \cdot \vec{\lambda}  }{m+ \lambda^0} 
							  + k^0 \frac{ \vec{\lambda}^2 }{ (m+ \lambda^0)^2 } 			\right]
			 \right\rbrace  \mathbb{I}_2 \notag \\
 =&	\frac{m+\lambda^0 }{2\varepsilon_{\rm eff} } 	\left\lbrace
			  1 +  \frac{ (\lambda^0)^2 -m^2  }{(m+ \lambda^0)^2} 
			 -   \frac{k^0\alpha }{\msf{\pi} \cdot \msf{k} }
							\left[  k^0 -  \frac{ 2\vec{k} \cdot \vec{\lambda}  }{m+ \lambda^0} 
							  + k^0 \frac{ (\lambda^0)^2 -m^2 }{ (m+ \lambda^0)^2 } 			\right]
			 \right\rbrace \mathbb{I}_2 \notag \\
 =&	 \mathbb{I}_2,
 \label{eq:convection1}
\end{align}
where $\msf{\lambda} \cdot \msf{\lambda} = m^2$ from \Eq{eq:dispersion_2}. Also notice that $\msf{\lambda} \cdot \msf{k} = \msf{\pi} \cdot \msf{k}$ from \Eqs{eq:k_dispersion} and \eq{eq:lambda}. Similarly,
\begin{align}
\Psi^\dag \gamma^0  \vec{\Gamma}  \Psi  =&  \frac{m+\lambda^0}{2 \varepsilon_{\rm eff}}  
 	\begin{pmatrix}
 		\mathbb{I}_2 &  \frac{ \vec{\sigma} \cdot \vec{\lambda} }{ m+ \lambda^0 } \\
	\end{pmatrix} 
	\left( \gamma^0 \vec{\gamma} - \vec{k} \frac{\alpha }{\msf{\pi} \cdot \msf{k} } \gamma^0 \sla{k} \right)
	\begin{pmatrix}
 		\mathbb{I}_2 \\ \frac{ \vec{\sigma} \cdot \vec{\lambda} }{m+ \lambda^0 } 
	\end{pmatrix} \notag \\
=& 
	\frac{m+\lambda^0}{2 \varepsilon_{\rm eff}}  
 	\begin{pmatrix}
 		\mathbb{I}_2 &  \frac{ \vec{\sigma} \cdot \vec{\lambda} }{m+\lambda^0  } \\
	\end{pmatrix} 
	\left[ 
		\begin{pmatrix}
			0 & \vec{\sigma}  \\ \vec{\sigma} & 0
		\end{pmatrix}
 		- \vec{k} \frac{\alpha }{\msf{\pi} \cdot \msf{k} }
		\begin{pmatrix}
		k^0 & - \vec{\sigma} \cdot \vec{k} \\ - \vec{\sigma} \cdot \vec{k}  & k^0 
		\end{pmatrix}
	\right]
	\begin{pmatrix}
 		\mathbb{I}_2 \\ \frac{ \vec{\sigma} \cdot \vec{\lambda} }{m+\lambda^0 } 
	\end{pmatrix} \notag \\
=& 
	\frac{m+\lambda^0}{2 \varepsilon_{\rm eff}}  
 	\begin{pmatrix}
		 \mathbb{I}_2 &  \frac{ \vec{\sigma} \cdot \vec{\lambda} }{m+\lambda^0  } \\
	\end{pmatrix} 
	\left[ 
		\begin{pmatrix}
			\frac{ \vec{\sigma} (\vec{\sigma} \cdot \vec{\lambda} ) }{m + \lambda^0 }   \\ 
			\vec{\sigma}
		\end{pmatrix}
 		- \vec{k} \frac{\alpha }{\msf{\pi} \cdot \msf{k} }
		\begin{pmatrix}
			k^0  \mathbb{I}_2 - \frac{ (\vec{\sigma} \cdot \vec{k}) (\vec{\sigma} \cdot \vec{\lambda} ) }{m+ \lambda^0 }   \\ 
			- \vec{\sigma} \cdot \vec{k} +k^0 \frac{ \vec{\sigma} \cdot \vec{\lambda} }{m+\lambda^0 } 
		\end{pmatrix}
	\right] \notag \\
=& 
	\frac{m+\lambda^0}{2 \varepsilon_{\rm eff}}  
	\left\lbrace 
		\frac{ \vec{\sigma} (\vec{\sigma} \cdot \vec{\lambda} ) + (\vec{\sigma} \cdot \vec{\lambda} ) \vec{\sigma}  }{m+\lambda^0 } -  \vec{k} \frac{\alpha }{\msf{\pi} \cdot \msf{k} }
		\left[ 
			k^0  - \frac{ 2\vec{k} \cdot \vec{\lambda}}{m+\lambda^0} + k^0 \frac{ \vec{\lambda}^2}{(m+\lambda^0)^2}
		\right]
	\right\rbrace 
	\mathbb{I}_2 \notag \\
=& 
	\frac{\vec{\lambda} -  \vec{k} \alpha }{\varepsilon_{\rm eff}}  
	\mathbb{I}_2 \notag \\
=& 
\frac{\vec{\pi}}{\varepsilon_{\rm eff}} \mathbb{I}_2.
\label{eq:convection2}
\end{align}
\end{widetext}
Hence, notice the following corollary of \Eqs{eq:convection1} and \eq{eq:convection2} that we will use below:
\begin{gather}
(\Psi^\dag \gamma^0 \Gamma^\mu \Psi) \pd_\mu
	=\mathbb{I}_2 \left(\pd_t + \frac{\vec{\pi}}{\varepsilon_{\rm eff}} \cdot \del \right)
	= \mathbb{I}_2 \mathrm{d}_t,
\label{eq:def_dt}
\end{gather}
where $\mathrm{d}_t$ is the same as defined in \Sec{sec:eigen_decomposition}. Substituting \Eq{eq:def_dt} into \Eq{eq:kinetic} leads to \Eq{eq:convection}.

\subsection{Expression for $\boldsymbol{\mc{U}}$ }
\label{app:reduced_potential}

An alternative representation of $\mc{U}$ in \Eq{eq:potential} is 
\begin{equation} \label{eq:U_aux}
\mc{U}  = -\mathrm{Im} \left[ \Psi^\dag \gamma^0  \Gamma^\mu  ( \pd_\mu \Psi  ) \right] ,
\end{equation}
where ``Im'' is short for the ``anti-Hermitian part of." To calculate $\pd_\mu \Psi$, let us consider $\Psi$ as a function
\begin{equation}
\Psi(t,\vec{x})=\Psi(\varepsilon_{\rm eff}(t,\vec{x}), \lambda^0(t,\vec{x}), \vec{\lambda}(t,\vec{x}) ).
\end{equation}
Notice that the contribution to \Eq{eq:U_aux} from the partial derivative with respect to $\varepsilon_{\rm eff}$ is zero. This is shown by using \Eqs{eq:projector} and \eq{eq:def_dt}:
\begin{align}
&\mathrm{Im} \left[  \Psi^\dag \gamma^0  \Gamma^\mu  ( \pd_{\varepsilon_{\rm eff}} \Psi  ) 
								\pd_\mu \varepsilon_{\rm eff}  \right] \notag \\
&~~~~~~~~~~~~	=	- \frac{1}{2}  \mathrm{Im} \left[  
									( \Psi^\dag \gamma^0  \Gamma^\mu \Psi  )   \pd_\mu \ln \varepsilon_{\rm eff}  \right] \notag \\
&~~~~~~~~~~~~	=	- \frac{1}{2}  \mathrm{Im} \left( 
									  \mathrm{d}_t
									  \ln \varepsilon_{\rm eff}   \right) \notag \\
&~~~~~~~~~~~~	=	 0 
\end{align}
since $\varepsilon_{\rm eff}$ is real. Then, $\mc{U} = -\mc{P}_t - \mc{P}_x-\mc{Q}_t-\mc{Q}_x$, where
\begin{align}
\mc{P}_t \doteq & 
		\mathrm{Im} \left[ \Psi^\dag \gamma^0  \Gamma^0  ( \pd_{\lambda_i} \Psi ) (\pd_t \lambda_i  ) \right],  \label{eq:dirac_aux_1} \\
\mc{P}_x \doteq &
		\mathrm{Im} \left[ \Psi^\dag \gamma^0  \vec{\Gamma}  ( \pd_{\lambda_i} \Psi ) \cdot (\del \lambda_i ) \right] , \\
\mc{Q}_t \doteq &
		\mathrm{Im} \left[ \Psi^\dag \gamma^0  \Gamma^0  ( \pd_{\lambda^0} \Psi ) (\pd_t \lambda^0 ) \right] , \\
\mc{Q}_x \doteq &
		\mathrm{Im} \left[ \Psi^\dag \gamma^0  \vec{\Gamma}  ( \pd_{\lambda^0} \Psi ) \cdot (\del \lambda^0 )  \right].
\label{eq:dirac_aux_4}
\end{align}

When substituting \Eqs{eq:Gamma}, \eq{eq:lambda}, \eq{eq:alpha}, and \eq{eq:projector} into $\mc{P}_t $, we obtain
\begin{widetext}
\begin{align}
\mc{P}_t =& \mathrm{Im} \left[ \Psi^\dag  \gamma^0  \Gamma^0 ( \pd_{\lambda_i} \Psi ) (\pd_t \lambda^i  ) \right] \notag \\
	=& \frac{m+\lambda^0}{2\varepsilon_{\rm eff}} \mathrm{Im} \left[ 
			\begin{pmatrix} \mathbb{I}_2 & \frac{\vec{\sigma}\cdot \vec{\lambda}}{m+\lambda^0 } \end{pmatrix}
			\gamma^0 \Gamma^0
 			\begin{pmatrix} 0 \\ \frac{\vec{\sigma}\cdot \pd_t \vec{\lambda}}{m+\lambda^0 } \end{pmatrix} \right] 
			\notag \\
	=& \frac{m+\lambda^0}{2\varepsilon_{\rm eff}} \mathrm{Im} \left\lbrace
			\begin{pmatrix} \mathbb{I}_2 & \frac{\vec{\sigma}\cdot \vec{\lambda}}{m+\lambda^0 } \end{pmatrix}
			\left[ \mathbb{I}_4  -  \frac{ k^0 \alpha }{\msf{\pi} \cdot \msf{k} }
							\begin{pmatrix}  k^0 & - \vec{\sigma} \cdot \vec{k} \\ - \vec{\sigma} \cdot \vec{k}  & k^0 \end{pmatrix}	 \right]
 			\begin{pmatrix} 0 \\ \frac{\vec{\sigma} \cdot \pd_t \vec{\lambda}}{m+\lambda^0 } \end{pmatrix} \right\rbrace
			\notag \\
	=& \frac{1}{2\varepsilon_{\rm eff}} \mathrm{Im} \left\lbrace
			\begin{pmatrix} \mathbb{I}_2 & \frac{\vec{\sigma}\cdot \vec{\lambda}}{m+\lambda^0 } \end{pmatrix}
			\left[ \begin{pmatrix} 0 \\ \vec{\sigma}\cdot \pd_t \vec{\lambda} \end{pmatrix} 
					 -  \frac{ k^0 \alpha }{\msf{\pi} \cdot \msf{k} }
							\begin{pmatrix}   -  (\vec{\sigma} \cdot \vec{k}) (\vec{\sigma}\cdot \pd_t \vec{\lambda})   \\ 
							\vec{\sigma}\cdot k^0 \pd_t \vec{\lambda} \end{pmatrix}	\right] \right\rbrace
			\notag \\
	=& \frac{1}{2\varepsilon_{\rm eff}} \mathrm{Im} \left[
			 \frac{(\vec{\sigma}\cdot \vec{\lambda})( \vec{\sigma}\cdot \pd_t \vec{\lambda} )  }{ m+\lambda^0 }   
					 + \frac{ k^0 \alpha }{\msf{\pi} \cdot \msf{k} } (\vec{\sigma} \cdot \vec{k}) (\vec{\sigma}\cdot \pd_t \vec{\lambda})
					 - \frac{ k^0 \alpha }{\msf{\pi} \cdot \msf{k} }  \frac{(\vec{\sigma}\cdot \vec{\lambda}) (\vec{\sigma}\cdot k^0 \pd_t \vec{\lambda} ) }{m+\lambda^0 } 
			\right] 			
			\notag \\
	=& \frac{1}{2\varepsilon_{\rm eff}} \vec{\sigma} \cdot  \left[
			 \frac{ \vec{\lambda} \times \pd_t \vec{\lambda}   }{ m+\lambda^0 }   
			 		- \frac{ k^0 \alpha }{\msf{\pi} \cdot \msf{k} }\frac{\vec{\lambda}\times k^0 \pd_t \vec{\lambda}}{m+\lambda^0}
					+\frac{  \alpha }{\msf{\pi} \cdot \msf{k} } \vec{k}\times k^0 \pd_t \vec{\lambda}
			\right] 				.
 \label{eq:res_1}
\end{align}
The next term, $\mc{P}_x$, is calculated similarly:
\begin{align}
\mc{P}_x =& \mathrm{Im} \left[ \Psi^\dag \gamma^0  \vec{\Gamma}  ( \pd_{\lambda_i} \Psi ) \cdot (\del \lambda^i )  \right] \notag \\
=& \frac{m+\lambda^0}{2\varepsilon_{\rm eff}} \mathrm{Im} \left\lbrace
		\begin{pmatrix} \mathbb{I}_2 & \frac{\vec{\sigma}\cdot \vec{\lambda}}{m+\lambda^0 } \end{pmatrix}
		\left[ \begin{pmatrix}  0 & \vec{\sigma}  \\ \vec{\sigma} & 0 \end{pmatrix} 
		- \vec{k} \frac{\alpha}{\msf{\pi} \cdot \msf{k}} 
				\begin{pmatrix} k^0 & - \vec{\sigma} \cdot \vec{k} \\ - \vec{\sigma} \cdot \vec{k}  & k^0  \end{pmatrix}  \right] \cdot
 		\begin{pmatrix} 0 \\ \frac{ \del ( \vec{\sigma} \cdot  \vec{\lambda}) }{m+\lambda^0 } \end{pmatrix}  \right\rbrace
 		\notag \\	
 =& \frac{1}{2\varepsilon_{\rm eff}} \mathrm{Im} \left\lbrace
		\begin{pmatrix} \mathbb{I}_2 & \frac{\vec{\sigma}\cdot \vec{\lambda}}{m+\lambda^0} \end{pmatrix}
		\left[
			\begin{pmatrix}  (\vec{\sigma} \cdot \del) ( \vec{\sigma} \cdot \vec{\lambda} )  \\ 0  \end{pmatrix}
		-\frac{\alpha}{\msf{\pi} \cdot \msf{k}}  
			\begin{pmatrix} -  (\vec{\sigma}\cdot \vec{k}) (\vec{k} \cdot \del) (\vec{\sigma} \cdot \vec{\lambda})   \\ 
				 k^0  (\vec{k} \cdot \del) (\vec{\sigma} \cdot \vec{\lambda})      \end{pmatrix}
		\right]
		 \right\rbrace
 		\notag \\	
  =& \frac{1}{2\varepsilon_{\rm eff}} \mathrm{Im} \left[		
		  	(\vec{\sigma} \cdot \del) ( \vec{\sigma} \cdot \vec{\lambda}) 
		 	+\frac{\alpha}{\msf{\pi} \cdot \msf{k}}  (\vec{\sigma}\cdot \vec{k}) (\vec{k} \cdot \del) (\vec{\sigma} \cdot \vec{\lambda}) 
			-\frac {k^0 \alpha}{\msf{\pi} \cdot \msf{k}}   \frac{  (\vec{\sigma}\cdot \vec{\lambda})  (\vec{k} \cdot \del) (\vec{\sigma} \cdot \vec{\lambda}) }{m+\lambda^0 } 
		 \right]
 		\notag \\	
   =& \frac{1}{2\varepsilon_{\rm eff}}  \vec{\sigma} \cdot   \left[		
		  \del \times \vec{\lambda}
		 	+\frac{        \alpha}{\msf{\pi} \cdot \msf{k}}    \vec{k} \times (\vec{k} \cdot \del)  \vec{\lambda}
			-\frac { k^0 \alpha}{\msf{\pi} \cdot \msf{k}} \frac{\vec{\lambda}  \times (\vec{k} \cdot \del)  \vec{\lambda} }{m+\lambda^0}      
		 \right]  .
\end{align}
Furthermore, the expressions for $\mc{Q}_t$ and $\mc{Q}_x$ are given by
\begin{align}
\mc{Q}_t =& \mathrm{Im} \left[ \Psi^\dag \gamma^0  \Gamma^0  ( \pd_{\lambda^0} \Psi ) (\pd_t \lambda^0 ) \right] \notag \\
	=& \frac{m+\lambda^0}{2\varepsilon_{\rm eff}} \mathrm{Im} \left\lbrace
			\begin{pmatrix} \mathbb{I}_2 & \frac{\vec{\sigma}\cdot \vec{\lambda}}{m+\lambda^0 } \end{pmatrix}
			\left[ \mathbb{I}_4  -  \frac{ k^0 \alpha }{\msf{\pi} \cdot \msf{k} }
							\begin{pmatrix}  k^0 & - \vec{\sigma} \cdot \vec{k} \\ - \vec{\sigma} \cdot \vec{k}  & k^0 \end{pmatrix}	 \right]
			\begin{pmatrix} 0 \\ - \frac{\vec{\sigma}\cdot  \vec{\lambda}}{(m+\lambda^0)^2 }\pd_t \lambda^0 \end{pmatrix} \right\rbrace
			\notag \\
	=& \frac{\pd_t \lambda^0}{2\varepsilon_{\rm eff}(m+\lambda^0)}  \mathrm{Im} \left\lbrace
			\begin{pmatrix} \mathbb{I}_2 & \frac{\vec{\sigma}\cdot \vec{\lambda}}{m+\lambda^0 } \end{pmatrix}
			\left[ 
				\begin{pmatrix} 0 \\ - \vec{\sigma}\cdot  \vec{\lambda}  \end{pmatrix}
				 -  \frac{ k^0 \alpha }{\msf{\pi} \cdot \msf{k} }
				\begin{pmatrix} 
					(\vec{\sigma}\cdot  \vec{k} ) (\vec{\sigma}\cdot  \vec{\lambda})  \\
					-k^0  \vec{\sigma} \cdot  \vec{\lambda}
				\end{pmatrix} 
			 \right]  
			 \right\rbrace
			\notag \\
	=& \frac{\pd_t \lambda^0}{2\varepsilon_{\rm eff}(m+\lambda^0)}  \mathrm{Im} \left[
				 - \frac{(\vec{\sigma}\cdot \vec{\lambda}) (\vec{\sigma}\cdot  \vec{\lambda} ) }{m+\lambda^0  }
				 -  \frac{ k^0 \alpha }{\msf{\pi} \cdot \msf{k} } (\vec{\sigma}\cdot  \vec{k} ) (\vec{\sigma}\cdot  \vec{\lambda}) 
				+  \frac{ (k^0)^2 \alpha }{\msf{\pi} \cdot \msf{k} } \frac{(\vec{\sigma} \cdot  \vec{\lambda}) (\vec{\sigma} \cdot  \vec{\lambda}) }{m+\lambda^0}
			 \right] 
			\notag \\
	=& - \frac{  k^0 \pd_t \lambda^0  }{ 2\varepsilon_{\rm eff} (m+\lambda^0) } 
				 \frac{ \alpha }{\msf{\pi} \cdot \msf{k} }   
				 \vec{\sigma} \cdot (\vec{k} \times \vec{\lambda}),
\end{align}
\begin{align}
\mc{Q}_x =& \mathrm{Im} \left[ \Psi^\dag \gamma^0  \vec{\Gamma}  ( \pd_{\lambda^0} \Psi ) \cdot (\del \lambda^0 ) \right] \notag \\
 =& \frac{m+\lambda^0}{2\varepsilon_{\rm eff}} \mathrm{Im} \left\lbrace
		\begin{pmatrix} \mathbb{I}_2 & \frac{\vec{\sigma}\cdot \vec{\lambda}}{m+\lambda^0 } \end{pmatrix}
		\left[ \begin{pmatrix}  0 & \vec{\sigma}  \\ \vec{\sigma} & 0 \end{pmatrix} 
		- \vec{k} \frac{\alpha}{\msf{\pi} \cdot \msf{k}} 
				\begin{pmatrix} k^0 & - \vec{\sigma} \cdot \vec{k} \\ - \vec{\sigma} \cdot \vec{k}  & k^0  \end{pmatrix}  
		\right] \cdot
 		\begin{pmatrix} 0 \\ - \frac{\vec{\sigma}\cdot  \vec{\lambda}}{(m+\lambda^0)^2 } \del\lambda^0 \end{pmatrix} 
 		\right\rbrace
 		\notag \\	
  =& \frac{1}{2\varepsilon_{\rm eff}(m+\lambda^0) } \mathrm{Im} \left\lbrace
		\begin{pmatrix} \mathbb{I}_2 & \frac{\vec{\sigma}\cdot \vec{\lambda}}{m+\lambda^0 } \end{pmatrix}
		\left[
			\begin{pmatrix}  - (\vec{\sigma} \cdot \del \lambda^0 ) ( \vec{\sigma} \cdot \vec{\lambda} )  \\ 0  \end{pmatrix}
		- (\vec{k} \cdot \del \lambda^0 ) \frac{\alpha}{\msf{\pi} \cdot \msf{k}}  
			\begin{pmatrix}  (\vec{\sigma}\cdot \vec{k} ) (\vec{\sigma} \cdot \vec{\lambda} )  \\ 
				- k^0   (\vec{\sigma} \cdot  \vec{\lambda} )    \end{pmatrix}
		\right]
		\right\rbrace
 		\notag \\	
   =& \frac{1}{2\varepsilon_{\rm eff}(m+\lambda^0) } \mathrm{Im} \left\lbrace
			 -  (\vec{\sigma} \cdot \del \lambda^0 ) ( \vec{\sigma} \cdot \vec{\lambda} )
		- (\vec{k} \cdot \del \lambda^0 ) \frac{\alpha}{\msf{\pi} \cdot \msf{k}}  
		\left[
			      (\vec{\sigma}\cdot \vec{k} ) (\vec{\sigma} \cdot \vec{\lambda} ) 
				- k^0   \frac{  (\vec{\sigma}\cdot \vec{\lambda})  (\vec{\sigma} \cdot \vec{\lambda} ) }{ m+\lambda^0 }    
		\right]
		\right\rbrace
 		\notag \\	
   =& \frac{1}{2\varepsilon_{\rm eff} (m+\lambda^0) } \vec{\sigma} \cdot \left[
		 \vec{\lambda} \times \del \lambda^0 
		-  \frac{\alpha}{\msf{\pi} \cdot \msf{k}}  ( \vec{k} \times \vec{\lambda}  ) (\vec{k} \cdot \del \lambda^0 ) 
		\right] .
\label{eq:res_4}
 \end{align}
Substituting \Eqs{eq:res_1}-\eq{eq:res_4} leads to
\begin{equation}
\mc{U}= - \frac{1}{2 \varepsilon_{\rm eff} } \vec{\sigma} \cdot \left[
			\del \times \vec{\lambda}
			+ \frac{\vec{\lambda} \times (  \del \lambda ^0	+	\pd_t \vec{\lambda}  ) }{m+\lambda^0} 
			+ \frac{\alpha}{\msf{\pi} \cdot \msf{k} } \vec{k}\times (k^\mu \pd_\mu ) \vec{\lambda}
			-  \frac{\alpha}{\msf{\pi} \cdot \msf{k} } \frac{ k^0 \vec{\lambda}\times  (k^\mu \pd_\mu ) \vec{\lambda}
						+  (\vec{k} \times \vec{\lambda}) (k^\mu \pd_\mu ) \lambda^0}{ m+\lambda^0}  \right].
\label{eq:U_all_terms}
\end{equation}

Equation \eq{eq:U_all_terms} can be simplified as follows. The first term can be rewritten as
\begin{align}
\del \times \vec{\lambda} =& \del \times ( \vec{\pi} +\vec{k}\alpha ) 
			= \del \times ( \del \theta -q\vec{A}_{\rm bg} +\vec{k} \alpha )
			= -q \vec{B}_{\rm bg} - \vec{k} \times \del \alpha
\label{eq:partial_I}
\end{align}
since $\del \times \vec{k} = \del \times \del \Theta =0$. Moreover, we note that $\pd^2_{\mu\nu}  \theta = \pd^2_{\nu\mu}\theta$. Hence,
\begin{align}
\del \lambda^0 =& \del \pi^0 + \alpha \del k^0 + k^0 \del \alpha \notag \\
		      \simeq & -\del (\pd_t \theta  + qV_{\rm bg} ) - \alpha \pd_t \vec{k}	+k^0 \del \alpha \notag \\
					=&  -\pd_t (\del \theta-q \vec{A}_{\rm bg} ) -q ( \del V_{\rm bg}  + \pd_t \vec{A}_{\rm bg} ) 
					- \alpha \pd_t \vec{k}	+k^0 \del \alpha \notag \\
					=&  -\pd_t \vec{\lambda} +q \vec{E}_{\rm bg} +	k^0 \del \alpha	+ \vec{k}	\pd_t \alpha .
\end{align}
Similarly, the numerator of the last term simplifies to

\begin{equation}
 	k^0 \vec{\lambda}			\times  (k^\mu \pd_\mu ) \vec{\lambda}	
 						+  (\vec{k} \times \vec{\lambda}) (k^\mu \pd_\mu ) \lambda^0 
			=	\vec{\lambda}	\times \left[	k^0 (k^\mu \pd_\mu )	\vec{\pi} 	-	\vec{k}	(k^\mu \pd_\mu ) \pi^0 	\right],
\label{eq:partial_IV}
\end{equation}
where
\begin{align}
(k^\mu \pd_\mu )\vec{k} =	&	k^0 \pd_t \vec{k} + (\vec{k}\cdot \del ) \vec{k}
						=		k^0 \del \pd_t \Theta	+ k^i	\del	\pd_i	\Theta	
						=	 -k^0 \del  k^0		+ k^i 	\del k_i		
						=	 - \del [ (k^0)^2 - \vec{k}^2 ]/2	
						=		0
\end{align}
and $(k^\mu \pd_\mu )k^0 	= 0$.
%
%
Here we used \Eq{eq:k_dispersion}. By substituting \Eqs{eq:partial_I}-\eq{eq:partial_IV} and explicitly showing the derivatives of $\alpha$, we obtain
\begin{align}
\mc{U}= &  \frac{q}{2 \varepsilon_{\rm eff} } \vec{\sigma} \cdot \left(
						\vec{B}_{\rm bg} - \frac{\vec{\lambda}\times  \vec{E}_{\rm bg}  }{m+\lambda^0}   
						\right) 
			 + \frac{q^2 }{4 \varepsilon_{\rm eff} \pik } \vec{\sigma} \cdot \left[
						 \vec{k}\times  \del \avg{ \msf{A}_{\rm osc}^2 } 
						- \frac{(\vec{\lambda} \times \vec{k} ) \pd_t \avg{ \msf{A}_{\rm osc}^2} }{m+\lambda^0}
						- \frac{ k^0   \vec{\lambda} \times \del \avg{ \msf{A}_{\rm osc}^2}  }{m+\lambda^0} 
						\right] \notag \\
			& + \frac{\alpha }{2 \varepsilon_{\rm eff}  \pik} \vec{\sigma} \cdot 
				\left(	\frac{k^0 \vec{\lambda}}{ m+\lambda^0} - \vec{k}	\right)	
						\times \left[ 	\del \pik +  (k^\mu \pd_\mu) \vec{\pi} 	\right] 
			  - \frac{\alpha }{2 \varepsilon_{\rm eff}  \pik}	\frac{\vec{\sigma} \cdot (\vec{\lambda}\times \vec{k}) }{m+\lambda^0} \left[ 
						(k^\mu \pd_\mu) \pi^0 - \pd_t \pik  
						\right] .
\label{eq:U_aux_II}
\end{align}
We can simplify the last two lines of \Eq{eq:U_aux_II} with
\begin{align}
 \del \pik  	+ (k^\mu \pd_\mu) \vec{\pi} 
 		 =		&	\del ( \pi^0 k^0	- \vec{k}\cdot \vec{\pi} ) 
 		 				+ k^0\pd_t \vec{\pi} + (\vec{k}\cdot \del) \vec{\pi}	\notag \\
 	\simeq	&	k^0 q \vec{E}_{\rm bg} + 		\pi^0 \del k^0		+	(\vec{k}\cdot \del ) \vec{\pi} 
 						- \del ( \vec{k}\cdot \vec{\pi} ) \notag \\
 		=		&	k^0 q \vec{E}_{\rm bg} 	- 	\pi^0 \del \pd_t \Theta	+	(\vec{k}\cdot \del ) \vec{\pi} 	
 					-\pi^i	\del \pd_i \Theta 	 - k_i	\del \pi^i  \notag \\
  		=		&	k^0 q \vec{E}_{\rm bg} 	- 	\pi^0 \pd_t \vec{k}	-(\vec{\pi}	\cdot 		\del 	) \vec{k}		
 					+	(\vec{k}\cdot \del ) \vec{\pi} 	- k_i	\del \pi^i  \notag \\
		= &	k^0 q \vec{E}_{\rm bg} -(\msf{\pi}^\mu \pd_\mu) \vec{k}		- \vec{k} \times (\del \times \vec{\pi} )	\notag	\\
		= &	k^0 q \vec{E}_{\rm bg} -(\msf{\pi}^\mu \pd_\mu) \vec{k}		- \vec{k} \times [\del \times (\del \theta -q\vec{A}_{\rm bg} )]		\notag	\\
		= &	k^0 q \vec{E}_{\rm bg} +	\vec{k}	\times q\vec{B}_{\rm bg}	-	(\msf{\pi}^\mu \pd_\mu) \vec{k}	,	\\
\notag 	\\
(k^\mu \pd_\mu) 	\pi^0 - \pd_t \pik  		
		=	&	k^0 \pd_t \pi^0 + \vec{k}\cdot \del \pi^0 - \pd_t (\pi^0 k^0		- \vec{\pi}\cdot \vec{k})  \notag \\
		=	&		\vec{k}\cdot \del  \pi^0 +  \pd_t (\vec{\pi}\cdot \vec{k})   -	\pi^0 \pd_t k^0			\notag \\
 \simeq 	&		 \vec{k}\cdot (-\pd_t \vec{\pi} +q \vec{E}_{\rm bg} ) +  \pd_t (\vec{\pi}\cdot \vec{k})   -	\pi^0 \pd_t k^0 \notag \\
		=	&		  q \vec{k}\cdot  \vec{E}_{\rm bg} -	\pi^0 \pd_t k^0 +\vec{\pi} \cdot \pd_t	\vec{k} 		\notag		\\
		=	&		  q \vec{k}\cdot  \vec{E}_{\rm bg} -\pi^0 \pd_t k^0 +\vec{\pi} \cdot \pd_t		\del	\Theta 		\notag		\\
		=	&		  q \vec{k}\cdot  \vec{E}_{\rm bg} -	\pi^0 \pd_t k^0 - (\vec{\pi} \cdot \del ) k^0 		\notag		\\
		=	&		  q \vec{k}\cdot  \vec{E}_{\rm bg} -	(\msf{\pi}^\mu \pd_\mu) k^0.
\end{align}
Hence, we obtain \Eqs{eq:potential_II} and \eq{eq:omega_eff}.

\end{widetext}


\begin{thebibliography}{49}
\expandafter\ifx\csname natexlab\endcsname\relax\def\natexlab#1{#1}\fi
\expandafter\ifx\csname bibnamefont\endcsname\relax
  \def\bibnamefont#1{#1}\fi
\expandafter\ifx\csname bibfnamefont\endcsname\relax
  \def\bibfnamefont#1{#1}\fi
\expandafter\ifx\csname citenamefont\endcsname\relax
  \def\citenamefont#1{#1}\fi
\expandafter\ifx\csname url\endcsname\relax
  \def\url#1{\texttt{#1}}\fi
\expandafter\ifx\csname urlprefix\endcsname\relax\def\urlprefix{URL }\fi
\providecommand{\bibinfo}[2]{#2}
\providecommand{\eprint}[2][]{\url{#2}}

\bibitem[{\citenamefont{Melrose}(2008)}]{Melrose:2008bw}
\bibinfo{author}{\bibfnamefont{D.~B.} \bibnamefont{Melrose}},
  \emph{\bibinfo{title}{{Quantum Plasmadynamics. Unmagnetized Plasmas}}}
  (\bibinfo{publisher}{Springer}, \bibinfo{address}{New York},
  \bibinfo{year}{2008}).

\bibitem[{\citenamefont{Brodin et~al.}(2008{\natexlab{a}})\citenamefont{Brodin,
  Marklund, and Manfredi}}]{Brodin:2008cl}
\bibinfo{author}{\bibfnamefont{G.}~\bibnamefont{Brodin}},
  \bibinfo{author}{\bibfnamefont{M.}~\bibnamefont{Marklund}}, \bibnamefont{and}
  \bibinfo{author}{\bibfnamefont{G.}~\bibnamefont{Manfredi}}, ``{Quantum plasma
  effects in the classical regime},'' \bibinfo{journal}{Phys. Rev. Lett.}
  \textbf{\bibinfo{volume}{100}}, \bibinfo{pages}{175001}
  (\bibinfo{year}{2008}{\natexlab{a}}).

\bibitem[{\citenamefont{Takabayasi}(1955)}]{Takabayasi:1955iy}
\bibinfo{author}{\bibfnamefont{T.}~\bibnamefont{Takabayasi}}, ``{The vector
  representation of spinning particle in the quantum theory. I},''
  \bibinfo{journal}{Prog. Theo. Phys.} \textbf{\bibinfo{volume}{14}},
  \bibinfo{pages}{283} (\bibinfo{year}{1955}).

\bibitem[{\citenamefont{Takabayasi}(1957)}]{Takabayasi:1957tz}
\bibinfo{author}{\bibfnamefont{T.}~\bibnamefont{Takabayasi}}, ``{Relativistic
  hydrodynamics of the Dirac matter},'' \bibinfo{journal}{Prog. Theo. Phys.
  Suppl.} \textbf{\bibinfo{volume}{4}}, \bibinfo{pages}{1}
  (\bibinfo{year}{1957}).

\bibitem[{\citenamefont{Marklund and Brodin}(2007)}]{Marklund:2007cv}
\bibinfo{author}{\bibfnamefont{M.}~\bibnamefont{Marklund}} \bibnamefont{and}
  \bibinfo{author}{\bibfnamefont{G.}~\bibnamefont{Brodin}}, ``{Dynamics of
  spin-1/2 quantum plasmas},'' \bibinfo{journal}{Phys. Rev. Lett.}
  \textbf{\bibinfo{volume}{98}}, \bibinfo{pages}{025001}
  (\bibinfo{year}{2007}).

\bibitem[{\citenamefont{Brodin et~al.}(2008{\natexlab{b}})\citenamefont{Brodin,
  Marklund, Zamanian, Ericsson, and Mana}}]{Brodin:2008er}
\bibinfo{author}{\bibfnamefont{G.}~\bibnamefont{Brodin}},
  \bibinfo{author}{\bibfnamefont{M.}~\bibnamefont{Marklund}},
  \bibinfo{author}{\bibfnamefont{J.}~\bibnamefont{Zamanian}},
  \bibinfo{author}{\bibfnamefont{{\AA}.}~\bibnamefont{Ericsson}},
  \bibnamefont{and} \bibinfo{author}{\bibfnamefont{P.~L.} \bibnamefont{Mana}},
  ``{Effects of the g factor in semiclassical kinetic plasma theory},''
  \bibinfo{journal}{Phys. Rev. Lett.} \textbf{\bibinfo{volume}{101}},
  \bibinfo{pages}{245002} (\bibinfo{year}{2008}{\natexlab{b}}).

\bibitem[{\citenamefont{Brodin et~al.}(2011)\citenamefont{Brodin, Marklund,
  Zamanian, and Stefan}}]{Brodin:2011kf}
\bibinfo{author}{\bibfnamefont{G.}~\bibnamefont{Brodin}},
  \bibinfo{author}{\bibfnamefont{M.}~\bibnamefont{Marklund}},
  \bibinfo{author}{\bibfnamefont{J.}~\bibnamefont{Zamanian}}, \bibnamefont{and}
  \bibinfo{author}{\bibfnamefont{M.}~\bibnamefont{Stefan}}, ``{Spin and
  magnetization effects in plasmas},'' \bibinfo{journal}{Plasma Phys. Control.
  Fusion} \textbf{\bibinfo{volume}{53}}, \bibinfo{pages}{074013}
  (\bibinfo{year}{2011}).

\bibitem[{\citenamefont{Stefan and Brodin}(2013)}]{Stefan:2013hr}
\bibinfo{author}{\bibfnamefont{M.}~\bibnamefont{Stefan}} \bibnamefont{and}
  \bibinfo{author}{\bibfnamefont{G.}~\bibnamefont{Brodin}}, ``{Linear and
  nonlinear wave propagation in weakly relativistic quantum plasmas},''
  \bibinfo{journal}{Phys. Plasmas} \textbf{\bibinfo{volume}{20}},
  \bibinfo{pages}{012114} (\bibinfo{year}{2013}).

\bibitem[{\citenamefont{Dixit et~al.}(2013)\citenamefont{Dixit, Hinschberger,
  Zamanian, Manfredi, and Hervieux}}]{Dixit:2013ff}
\bibinfo{author}{\bibfnamefont{A.}~\bibnamefont{Dixit}},
  \bibinfo{author}{\bibfnamefont{Y.}~\bibnamefont{Hinschberger}},
  \bibinfo{author}{\bibfnamefont{J.}~\bibnamefont{Zamanian}},
  \bibinfo{author}{\bibfnamefont{G.}~\bibnamefont{Manfredi}}, \bibnamefont{and}
  \bibinfo{author}{\bibfnamefont{P.-A.} \bibnamefont{Hervieux}}, ``{Lagrangian
  approach to the semirelativistic electron dynamics in the mean-field
  approximation},'' \bibinfo{journal}{Phys. Rev. A}
  \textbf{\bibinfo{volume}{88}}, \bibinfo{pages}{032117}
  (\bibinfo{year}{2013}).

\bibitem[{\citenamefont{Morandi et~al.}(2014)\citenamefont{Morandi, Zamanian,
  Manfredi, and Hervieux}}]{Morandi:2014jm}
\bibinfo{author}{\bibfnamefont{O.}~\bibnamefont{Morandi}},
  \bibinfo{author}{\bibfnamefont{J.}~\bibnamefont{Zamanian}},
  \bibinfo{author}{\bibfnamefont{G.}~\bibnamefont{Manfredi}}, \bibnamefont{and}
  \bibinfo{author}{\bibfnamefont{P.-A.} \bibnamefont{Hervieux}},
  ``{Quantum-relativistic hydrodynamic model for a spin-polarized electron gas
  interacting with light},'' \bibinfo{journal}{Phys. Rev. E}
  \textbf{\bibinfo{volume}{90}}, \bibinfo{pages}{013103}
  (\bibinfo{year}{2014}).

\bibitem[{\citenamefont{Andreev and
  Kuz'menkov}(2014{\natexlab{a}})}]{Andreev:1746689}
\bibinfo{author}{\bibfnamefont{P.~A.} \bibnamefont{Andreev}} \bibnamefont{and}
  \bibinfo{author}{\bibfnamefont{L.~S.} \bibnamefont{Kuz'menkov}},
  ``{Many-particle quantum hydrodynamics: Basic principles and fundamental
  applications},'' \bibinfo{journal}{arXiv}
  (\bibinfo{year}{2014}{\natexlab{a}}), \eprint{1407.7770}.

\bibitem[{\citenamefont{Andreev}(2015)}]{Andreev:2015if}
\bibinfo{author}{\bibfnamefont{P.~A.} \bibnamefont{Andreev}}, ``{Separated
  spin-up and spin-down quantum hydrodynamics of degenerated electrons:
  Spin-electron acoustic wave appearance},'' \bibinfo{journal}{Phys. Rev. E}
  \textbf{\bibinfo{volume}{91}}, \bibinfo{pages}{033111}
  (\bibinfo{year}{2015}).

\bibitem[{\citenamefont{Andreev and
  Kuz'menkov}(2014{\natexlab{b}})}]{Andreev:2014uw}
\bibinfo{author}{\bibfnamefont{P.~A.} \bibnamefont{Andreev}} \bibnamefont{and}
  \bibinfo{author}{\bibfnamefont{L.~S.} \bibnamefont{Kuz'menkov}}, ``{Oblique
  propagation of longitudinal waves in magnetized spin-1/2 plasmas: independent
  evolution of spin-up and spin-down electrons},'' \bibinfo{journal}{arXiv}
  (\bibinfo{year}{2014}{\natexlab{b}}), \eprint{1406.6252}.

\bibitem[{\citenamefont{Boot and Harvie}(1957)}]{Boot:1957um}
\bibinfo{author}{\bibfnamefont{H.~A.~H.} \bibnamefont{Boot}} \bibnamefont{and}
  \bibinfo{author}{\bibfnamefont{R.~B. R.-S.} \bibnamefont{Harvie}}, ``{Charged
  particles in a non-uniform radio-frequency field},''
  \bibinfo{journal}{Nature} \textbf{\bibinfo{volume}{180}},
  \bibinfo{pages}{1187} (\bibinfo{year}{1957}).

\bibitem[{\citenamefont{Gaponov and Miller}(1958)}]{Gaponov:1958un}
\bibinfo{author}{\bibfnamefont{A.~V.} \bibnamefont{Gaponov}} \bibnamefont{and}
  \bibinfo{author}{\bibfnamefont{M.~A.} \bibnamefont{Miller}},
  ``{Quasipotential theory},'' \bibinfo{journal}{Sov. Phys. JETP}
  \textbf{\bibinfo{volume}{7}}, \bibinfo{pages}{168} (\bibinfo{year}{1958}).

\bibitem[{\citenamefont{Cary and Kaufman}(1977)}]{Cary:1977ud}
\bibinfo{author}{\bibfnamefont{J.~R.} \bibnamefont{Cary}} \bibnamefont{and}
  \bibinfo{author}{\bibfnamefont{A.~N.} \bibnamefont{Kaufman}},
  ``{Ponderomotive force and linear susceptibility in Vlasov plasma},''
  \bibinfo{journal}{Phys. Rev. Lett.} \textbf{\bibinfo{volume}{39}},
  \bibinfo{pages}{402} (\bibinfo{year}{1977}).

\bibitem[{\citenamefont{Brodin et~al.}(2010)\citenamefont{Brodin, Misra, and
  Marklund}}]{Brodin:2010bc}
\bibinfo{author}{\bibfnamefont{G.}~\bibnamefont{Brodin}},
  \bibinfo{author}{\bibfnamefont{A.~P.} \bibnamefont{Misra}}, \bibnamefont{and}
  \bibinfo{author}{\bibfnamefont{M.}~\bibnamefont{Marklund}}, ``{Spin
  contribution to the ponderomotive force in a plasma},''
  \bibinfo{journal}{Phys. Rev. Lett.} \textbf{\bibinfo{volume}{105}},
  \bibinfo{pages}{105004} (\bibinfo{year}{2010}).

\bibitem[{\citenamefont{Stefan et~al.}(2011)\citenamefont{Stefan, Zamanian,
  Brodin, Misra, and Marklund}}]{Stefan:2011cq}
\bibinfo{author}{\bibfnamefont{M.}~\bibnamefont{Stefan}},
  \bibinfo{author}{\bibfnamefont{J.}~\bibnamefont{Zamanian}},
  \bibinfo{author}{\bibfnamefont{G.}~\bibnamefont{Brodin}},
  \bibinfo{author}{\bibfnamefont{A.~P.} \bibnamefont{Misra}}, \bibnamefont{and}
  \bibinfo{author}{\bibfnamefont{M.}~\bibnamefont{Marklund}}, ``{Ponderomotive
  force due to the intrinsic spin in extended fluid and kinetic models},''
  \bibinfo{journal}{Phys. Rev. E} \textbf{\bibinfo{volume}{83}},
  \bibinfo{pages}{036410} (\bibinfo{year}{2011}).

\bibitem[{\citenamefont{Akhiezer and Polovin}(1956)}]{Akhiezer:1956wb}
\bibinfo{author}{\bibfnamefont{A.~I.} \bibnamefont{Akhiezer}} \bibnamefont{and}
  \bibinfo{author}{\bibfnamefont{R.~V.} \bibnamefont{Polovin}}, ``{Theory of
  wave motion of an electron plasma},'' \bibinfo{journal}{Sov. Phys. JETP}
  \textbf{\bibinfo{volume}{3}}, \bibinfo{pages}{696} (\bibinfo{year}{1956}).

\bibitem[{\citenamefont{Kibble}(1966{\natexlab{a}})}]{Kibble:1966fj}
\bibinfo{author}{\bibfnamefont{T.~W.~B.} \bibnamefont{Kibble}}, ``{Mutual
  refraction of electrons and photons},'' \bibinfo{journal}{Phys. Rev.}
  \textbf{\bibinfo{volume}{150}}, \bibinfo{pages}{1060}
  (\bibinfo{year}{1966}{\natexlab{a}}).

\bibitem[{\citenamefont{Kibble}(1966{\natexlab{b}})}]{Kibble:1966eb}
\bibinfo{author}{\bibfnamefont{T.~W.~B.} \bibnamefont{Kibble}}, ``{Refraction
  of electron beams by intense electromagnetic waves},''
  \bibinfo{journal}{Phys. Rev. Lett.} \textbf{\bibinfo{volume}{16}},
  \bibinfo{pages}{1054} (\bibinfo{year}{1966}{\natexlab{b}}).

\bibitem[{\citenamefont{Raicher et~al.}(2014)\citenamefont{Raicher, Eliezer,
  and Zigler}}]{Raicher:2014gj}
\bibinfo{author}{\bibfnamefont{E.}~\bibnamefont{Raicher}},
  \bibinfo{author}{\bibfnamefont{S.}~\bibnamefont{Eliezer}}, \bibnamefont{and}
  \bibinfo{author}{\bibfnamefont{A.}~\bibnamefont{Zigler}}, ``{The Lagrangian
  formulation of strong-field quantum electrodynamics in a plasma},''
  \bibinfo{journal}{Phys. Plasmas} \textbf{\bibinfo{volume}{21}},
  \bibinfo{pages}{053103} (\bibinfo{year}{2014}).

\bibitem[{foo({\natexlab{a}})}]{foot:dodin}
\bibinfo{note}{Under certain conditions, the effective-mass theory is also
  extendable to interactions in the presence of arbitrarily strong large-scale
  magnetic fields [I. Y. Dodin and N. J. Fisch, Phys. Rev. E {\bf 77}, 036402
  (2008)] and plasmas with small yet nonzero background density [V. I. Geyko,
  G. M. Fraiman, I. Y. Dodin, and N. J. Fisch, Phys. Rev. E \textbf{80}, 036404
  (2009)].}

\bibitem[{\citenamefont{Dirac}(1928)}]{Dirac:1928dh}
\bibinfo{author}{\bibfnamefont{P.~A.~M.} \bibnamefont{Dirac}}, ``{The quantum
  theory of electron},'' \bibinfo{journal}{Proc. R. Soc. Lond. A.}
  \textbf{\bibinfo{volume}{117}}, \bibinfo{pages}{610} (\bibinfo{year}{1928}).

\bibitem[{\citenamefont{Tracy et~al.}(2014)\citenamefont{Tracy, Brizard,
  Richardson, and Kaufman}}]{Tracy:2014to}
\bibinfo{author}{\bibfnamefont{E.~R.} \bibnamefont{Tracy}},
  \bibinfo{author}{\bibfnamefont{A.~J.} \bibnamefont{Brizard}},
  \bibinfo{author}{\bibfnamefont{A.~S.} \bibnamefont{Richardson}},
  \bibnamefont{and} \bibinfo{author}{\bibfnamefont{A.~N.}
  \bibnamefont{Kaufman}}, \emph{\bibinfo{title}{{Ray Tracing and Beyond: Phase
  Space Methods in Plasma Wave Theory}}} (\bibinfo{publisher}{Cambridge
  University Press}, \bibinfo{address}{New York}, \bibinfo{year}{2014}).

\bibitem[{\citenamefont{Bargmann et~al.}(1959)\citenamefont{Bargmann, Michel,
  and Telegdi}}]{Bargmann:1959us}
\bibinfo{author}{\bibfnamefont{V.}~\bibnamefont{Bargmann}},
  \bibinfo{author}{\bibfnamefont{L.}~\bibnamefont{Michel}}, \bibnamefont{and}
  \bibinfo{author}{\bibfnamefont{V.~L.} \bibnamefont{Telegdi}}, ``{Precession
  of the polarization of particles moving in a homogeneous electromagnetic
  field},'' \bibinfo{journal}{Phys. Rev. Lett.} \textbf{\bibinfo{volume}{2}},
  \bibinfo{pages}{435} (\bibinfo{year}{1959}).

\bibitem[{\citenamefont{Hairer et~al.}(2006)\citenamefont{Hairer, Lubich, and
  Wanner}}]{Hairer_2006}
\bibinfo{author}{\bibfnamefont{E.}~\bibnamefont{Hairer}},
  \bibinfo{author}{\bibfnamefont{C.}~\bibnamefont{Lubich}}, \bibnamefont{and}
  \bibinfo{author}{\bibfnamefont{G.}~\bibnamefont{Wanner}},
  \emph{\bibinfo{title}{{Geometric Numerical Integration}}}
  (\bibinfo{publisher}{Springer}, \bibinfo{year}{2006}).

\bibitem[{\citenamefont{McLachlan and Quispel}(2006)}]{McLachlan_2006}
\bibinfo{author}{\bibfnamefont{R.~I.} \bibnamefont{McLachlan}}
  \bibnamefont{and} \bibinfo{author}{\bibfnamefont{G.~R.~W.}
  \bibnamefont{Quispel}}, ``{Geometric integrators for ODEs},''
  \bibinfo{journal}{J. Phys. A: Math. Theor.} \textbf{\bibinfo{volume}{39}},
  \bibinfo{pages}{5251} (\bibinfo{year}{2006}).

\bibitem[{\citenamefont{Qin et~al.}(2015)\citenamefont{Qin, Liu, Xiao, Zhang,
  He, Wang, Sun, Burby, Ellison, and Zhou}}]{Qin_2015}
\bibinfo{author}{\bibfnamefont{H.}~\bibnamefont{Qin}},
  \bibinfo{author}{\bibfnamefont{J.}~\bibnamefont{Liu}},
  \bibinfo{author}{\bibfnamefont{J.}~\bibnamefont{Xiao}},
  \bibinfo{author}{\bibfnamefont{R.}~\bibnamefont{Zhang}},
  \bibinfo{author}{\bibfnamefont{Y.}~\bibnamefont{He}},
  \bibinfo{author}{\bibfnamefont{Y.}~\bibnamefont{Wang}},
  \bibinfo{author}{\bibfnamefont{Y.}~\bibnamefont{Sun}},
  \bibinfo{author}{\bibfnamefont{J.}~\bibnamefont{Burby}},
  \bibinfo{author}{\bibfnamefont{C.~L.} \bibnamefont{Ellison}},
  \bibnamefont{and} \bibinfo{author}{\bibfnamefont{Y.}~\bibnamefont{Zhou}},
  ``{Canonical symplectic particle-in-cell method for long-term large-scale
  simulations of the Vlasov-Maxwell equations},'' \bibinfo{journal}{Accepted to
  Nuclear Fusion}  (\bibinfo{year}{2015}).

\bibitem[{\citenamefont{Dodin}(2014)}]{Dodin:2014hw}
\bibinfo{author}{\bibfnamefont{I.~Y.} \bibnamefont{Dodin}}, ``{Geometric view
  on noneikonal waves},'' \bibinfo{journal}{Phys. Lett. A}
  \textbf{\bibinfo{volume}{378}}, \bibinfo{pages}{1598} (\bibinfo{year}{2014}).

\bibitem[{foo({\natexlab{b}})}]{foot:anomalous}
\bibinfo{note}{In this work, the small contribution of the anomalous magnetic
  moment term is neglected but could be included too, at least as a
  perturbation.}

\bibitem[{\citenamefont{Dodin and Fisch}(2012)}]{Dodin:2012hn}
\bibinfo{author}{\bibfnamefont{I.~Y.} \bibnamefont{Dodin}} \bibnamefont{and}
  \bibinfo{author}{\bibfnamefont{N.~J.} \bibnamefont{Fisch}}, ``{Axiomatic
  geometrical optics, Abraham-Minkowski controversy, and photon properties
  derived classically},'' \bibinfo{journal}{Phys. Rev. A}
  \textbf{\bibinfo{volume}{86}}, \bibinfo{pages}{053834}
  (\bibinfo{year}{2012}).

\bibitem[{\citenamefont{Ruiz and Dodin}(2015{\natexlab{a}})}]{Ruiz:2015hq}
\bibinfo{author}{\bibfnamefont{D.~E.} \bibnamefont{Ruiz}} \bibnamefont{and}
  \bibinfo{author}{\bibfnamefont{I.~Y.} \bibnamefont{Dodin}}, ``{Lagrangian
  geometrical optics of nonadiabatic vector waves and spin particles},''
  \bibinfo{journal}{Phys. Lett. A} \textbf{\bibinfo{volume}{379}},
  \bibinfo{pages}{2337} (\bibinfo{year}{2015}{\natexlab{a}}).

\bibitem[{\citenamefont{Ruiz and Dodin}(2015{\natexlab{b}})}]{Ruiz:2015bz}
\bibinfo{author}{\bibfnamefont{D.~E.} \bibnamefont{Ruiz}} \bibnamefont{and}
  \bibinfo{author}{\bibfnamefont{I.~Y.} \bibnamefont{Dodin}}, ``{On the
  correspondence between quantum and classical variational principles},''
  \bibinfo{journal}{Phys. Lett. A} \textbf{\bibinfo{volume}{379}},
  \bibinfo{pages}{2623} (\bibinfo{year}{2015}{\natexlab{b}}).

\bibitem[{\citenamefont{Whitham}(1965)}]{Whitham:1965kx}
\bibinfo{author}{\bibfnamefont{G.~B.} \bibnamefont{Whitham}}, ``{A general
  approach to linear and non-linear dispersive waves using a Lagrangian},''
  \bibinfo{journal}{J. Fluid Mech.} \textbf{\bibinfo{volume}{22}},
  \bibinfo{pages}{273} (\bibinfo{year}{1965}).

\bibitem[{\citenamefont{Whitham}(2011)}]{Whitham:2011kb}
\bibinfo{author}{\bibfnamefont{G.~B.} \bibnamefont{Whitham}},
  \emph{\bibinfo{title}{{Linear and Nonlinear Waves}}}
  (\bibinfo{publisher}{Wiley}, \bibinfo{address}{New York},
  \bibinfo{year}{2011}).

\bibitem[{\citenamefont{Yakovlev}(1966)}]{Yakovlev:1966tp}
\bibinfo{author}{\bibfnamefont{V.~P.} \bibnamefont{Yakovlev}},
  ``{Electron-positron pair production by a strong electromagnetic wave in the
  field of a nucleus},'' \bibinfo{journal}{Sov. Phys. JETP}
  \textbf{\bibinfo{volume}{22}}, \bibinfo{pages}{223} (\bibinfo{year}{1966}).

\bibitem[{\citenamefont{Thaller}(1992)}]{Thaller:1992th}
\bibinfo{author}{\bibfnamefont{B.}~\bibnamefont{Thaller}},
  \emph{\bibinfo{title}{{The Dirac Equation}}} (\bibinfo{publisher}{Springer},
  \bibinfo{address}{Berlin}, \bibinfo{year}{1992}).

\bibitem[{\citenamefont{Malka et~al.}(1997)\citenamefont{Malka, Lefebvre, and
  Miquel}}]{Malka:1997tc}
\bibinfo{author}{\bibfnamefont{G.}~\bibnamefont{Malka}},
  \bibinfo{author}{\bibfnamefont{E.}~\bibnamefont{Lefebvre}}, \bibnamefont{and}
  \bibinfo{author}{\bibfnamefont{J.~L.} \bibnamefont{Miquel}}, ``{Experimental
  observation of electrons accelerated in vacuum to relativistic energies by a
  high-intensity laser},'' \bibinfo{journal}{Phys. Rev. Lett.}
  \textbf{\bibinfo{volume}{78}}, \bibinfo{pages}{3314} (\bibinfo{year}{1997}).

\bibitem[{\citenamefont{Quesnel and Mora}(1998)}]{Quesnel:1998tj}
\bibinfo{author}{\bibfnamefont{B.}~\bibnamefont{Quesnel}} \bibnamefont{and}
  \bibinfo{author}{\bibfnamefont{P.}~\bibnamefont{Mora}}, ``{Theory and
  simulation of the interaction of ultra-intense laser pulses with electrons in
  vacuum},'' \bibinfo{journal}{Phys. Rev. E} \textbf{\bibinfo{volume}{58}},
  \bibinfo{pages}{3719} (\bibinfo{year}{1998}).

\bibitem[{\citenamefont{Mora and Antonsen~Jr.}(1997)}]{Mora:1997ku}
\bibinfo{author}{\bibfnamefont{P.}~\bibnamefont{Mora}} \bibnamefont{and}
  \bibinfo{author}{\bibfnamefont{T.~M.} \bibnamefont{Antonsen~Jr.}}, ``{Kinetic
  modeling of intense, short laser pulses propagating in tenuous plasmas},''
  \bibinfo{journal}{Phys. Plasmas} \textbf{\bibinfo{volume}{4}},
  \bibinfo{pages}{217} (\bibinfo{year}{1997}).

\bibitem[{\citenamefont{Dodin et~al.}(2003)\citenamefont{Dodin, Fisch, and
  Fraiman}}]{Dodin:2003hz}
\bibinfo{author}{\bibfnamefont{I.~Y.} \bibnamefont{Dodin}},
  \bibinfo{author}{\bibfnamefont{N.~J.} \bibnamefont{Fisch}}, \bibnamefont{and}
  \bibinfo{author}{\bibfnamefont{G.~M.} \bibnamefont{Fraiman}}, ``{Drift
  Lagrangian for a relativistic particle in an intense laser field},''
  \bibinfo{journal}{JETP Lett.} \textbf{\bibinfo{volume}{78}},
  \bibinfo{pages}{202} (\bibinfo{year}{2003}).

\bibitem[{\citenamefont{Ruiz and Dodin}(2015{\natexlab{c}})}]{Ruiz:2015dv}
\bibinfo{author}{\bibfnamefont{D.~E.} \bibnamefont{Ruiz}} \bibnamefont{and}
  \bibinfo{author}{\bibfnamefont{I.~Y.} \bibnamefont{Dodin}},
  ``{First-principles variational formulation of polarization effects in
  geometrical optics},'' \bibinfo{journal}{Phys. Rev. A}
  \textbf{\bibinfo{volume}{92}}, \bibinfo{pages}{043805}
  (\bibinfo{year}{2015}{\natexlab{c}}).

\bibitem[{\citenamefont{Hayes}(1973)}]{Hayes:1973dt}
\bibinfo{author}{\bibfnamefont{W.~D.} \bibnamefont{Hayes}}, ``{Group velocity
  and nonlinear dispersive wave propagation},'' \bibinfo{journal}{Proc. R. Soc.
  Lond. A.} \textbf{\bibinfo{volume}{332}}, \bibinfo{pages}{199}
  (\bibinfo{year}{1973}).

\bibitem[{\citenamefont{Mora and Antonsen~Jr.}(1996)}]{Mora:1996jy}
\bibinfo{author}{\bibfnamefont{P.}~\bibnamefont{Mora}} \bibnamefont{and}
  \bibinfo{author}{\bibfnamefont{T.~M.} \bibnamefont{Antonsen~Jr.}},
  ``{Electron cavitation and acceleration in the wake of an ultraintense,
  self-focused laser pulse},'' \bibinfo{journal}{Phys. Rev. E}
  \textbf{\bibinfo{volume}{53}}, \bibinfo{pages}{R2068} (\bibinfo{year}{1996}).

\bibitem[{\citenamefont{Mora and Quesnel}(1998)}]{Mora:1998zz}
\bibinfo{author}{\bibfnamefont{P.}~\bibnamefont{Mora}} \bibnamefont{and}
  \bibinfo{author}{\bibfnamefont{B.}~\bibnamefont{Quesnel}}, ``{Comment on
  `Experimental observation of electrons accelerated in vacuum to relativistic
  energies by a high-intensity laser'},'' \bibinfo{journal}{Phys. Rev. Lett.}
  \textbf{\bibinfo{volume}{80}}, \bibinfo{pages}{1351} (\bibinfo{year}{1998}).

\bibitem[{\citenamefont{Volkov}(1935)}]{Volkov:1935vka}
\bibinfo{author}{\bibfnamefont{D.~M.} \bibnamefont{Volkov}}, ``{On a class of
  solutions of the Dirac equation},'' \bibinfo{journal}{Z. Phys.}
  \textbf{\bibinfo{volume}{94}}, \bibinfo{pages}{250} (\bibinfo{year}{1935}).

\bibitem[{\citenamefont{Bergou and Varro}(1980)}]{Vergou:1979}
\bibinfo{author}{\bibfnamefont{J.}~\bibnamefont{Bergou}} \bibnamefont{and}
  \bibinfo{author}{\bibfnamefont{S.}~\bibnamefont{Varro}}, ``{Wavefunctions of
  a free-electron in an external field and their application in intense field
  interactions: II. relativistic treatment},'' \bibinfo{journal}{J. Phys. A:
  Math. Gen.} \textbf{\bibinfo{volume}{13}}, \bibinfo{pages}{2823}
  (\bibinfo{year}{1980}).

\bibitem[{\citenamefont{Raicher and Eliezer}(2013)}]{Raicher:2013ik}
\bibinfo{author}{\bibfnamefont{E.}~\bibnamefont{Raicher}} \bibnamefont{and}
  \bibinfo{author}{\bibfnamefont{S.}~\bibnamefont{Eliezer}}, ``{Analytical
  solutions of the Dirac and the Klein-Gordon equations in plasma induced by
  high-intensity laser},'' \bibinfo{journal}{Phys. Rev. A}
  \textbf{\bibinfo{volume}{88}}, \bibinfo{pages}{022113}
  (\bibinfo{year}{2013}).

\end{thebibliography}
\end{document}